\documentclass[preprint2,times,tighten]{aastex6}
\usepackage{amsmath,amstext}
\usepackage[figure,figure*]{hypcap}
\usepackage{natbib}

\newcommand\arcpt{${{\lower3pt\hbox{$^{\prime\prime}$}}\atop{\raise4pt\hbox{.}}}$}

\def\arcsec{\ifmmode {''}\else $''$\fi}

\slugcomment{submitted to the {\it Astrophysical Journal} }

\shorttitle{SALT Young Stars}
\shortauthors{Riedel et al.}

\begin{document}

\title{Young Stars with SALT\thanks{based on observations made with the Southern African Large Telescope (SALT)}}

\author{Adric~R.~Riedel}

\affil {Department of Astronomy, California Institute of Technology, Pasadena, CA 91125}
\affil {Department of Astrophysics, The American Museum of Natural History, New York, NY, 10024}
\affil {Department of Engineering Science and Physics, The College of Staten Island, Staten Island, NY, 10314} 
\affil {Department of Physics and Astronomy, Hunter College, New York, NY, 10065} 

\email{arr@caltech.edu}

\author {Munazza~K.~Alam}

\affil {Department of Astronomy, Harvard University, Cambridge, MA 02138}
\affil {Department of Astrophysics, The American Museum of Natural History, New York, NY, 10024}
\affil {Department of Physics and Astronomy, Hunter College, New York, NY, 10065}

\author {Emily~L.~Rice}
\affil {Department of Engineering Science and Physics, The College of Staten Island, Staten Island, NY, 10314}
\affil {Department of Astrophysics, The American Museum of Natural History, New York, NY, 10024}
\affil {Physics Program, The Graduate Center, CUNY, New York, NY 10016}

\author {Kelle L. Cruz}
\affil {Department of Physics and Astronomy, Hunter College, City University of New York, New York, NY, 10065}
\affil {Department of Astrophysics, The American Museum of Natural History, New York, NY, 10024}
\affil {Physics Program, The Graduate Center, CUNY, New York, NY 10016}

\author {Todd J. Henry}
\affil {RECONS Institute, Chambersburg, PA.} 

\begin{abstract}

We present a spectroscopic and kinematic analysis of 79 nearby M dwarfs in 77 systems. All are low-proper-motion southern hemisphere objects and were identified in a nearby star survey with a demonstrated sensitivity to young stars. Using low-resolution optical spectroscopy from the Red Side Spectrograph (RSS) on the South African Large Telescope (SALT), we have determined radial velocities, H-alpha, Lithium 6708\AA, and Potassium 7699\AA~equivalent widths linked to age and activity, and spectral types for all our targets.  Combined with astrometric information from literature sources, we identify 44 young stars. Eighteen are previously known members of moving groups within 100 parsecs of the Sun. Twelve are new members, including one member of the TW Hydra moving group, one member of the 32 Orionis moving group, nine members of Tucana-Horologium, one member of Argus, and two new members of AB Doradus. We also find fourteen young star systems that are not members of any known groups. The remaining 33 star systems do not appear to be young. This appears to be evidence of a new population of nearby young stars not related to the known nearby young moving groups.


\end{abstract}

\section{Introduction}
\label{sec:intro}

The Nearby Young Moving Groups (NYMGs), loose associations of nearby stars between 5 ($\epsilon$ Cha, \citealt{Murphy2013}) and 525 Myr old ($\chi^1$ For, \citealt{Pohnl2010}), are thought to be remnants of small-scale star formation in the nearby Sco-Cen star forming region and represent the closest assemblages of pre-main-sequence stars and young planetary systems to the Sun. They are valuable targets for studying the formation of low mass stars, brown dwarfs, and planetary systems, because their proximity makes it easier to study fainter objects and companions at smaller separations.

A number of investigators have dedicated time to large-scale surveys for members of the nearby young moving groups (e.g. \citealt{Torres2000}, \citealt{Song2003}, \citealt{Shkolnik2009}, \citealt{Schlieder2010}, \citealt{Murphy2010}, \citealt{Rodriguez2011}, \citealt{Riedel2011}, \citealt{Malo2013}, \citealt{Rodriguez2013}, \citealt{Gagne2014a} and subsequent), and thanks to their efforts our samples of low-mass stars to study have been continually growing: we now know of over 650 M-dwarf and lower-mass members of the nearby young moving groups, identified by spectroscopic signs of low surface gravity, age-related lithium absorption, and kinematic matches to the groups.

The question remains, however: given the success of these programs at identifying NYMG members, how many more remain to be discovered? The RECONS\footnote{Research Consortium On Nearby Stars, \url{http://www.recons.org}} TINY proper MOtions (TINYMO) survey \citep{Riedel2012} delivered a number of new moving group members, many already published in \citet{Riedel2014}. Given that the survey has already yielded 26 young stars out of the 55 that have been followed up with astrometry, it is reasonable to assume that the survey may contain many other young stars.

The TINYMO survey contains proper motions measured from the SuperCOSMOS Sky Survey \citep{Hambly2001a}, but these are not sufficient to identify new nearby young stars. A study of moving group identification codes (\citealt{Riedel2016a}) demonstrates that it is impossible to be certain about memberships in a moving group based on proper motions alone; the addition of either radial velocity or parallax measurements (preferably both) dramatically increases the quality of kinematic membership assignments. 

Even so, kinematic memberships are only part of the puzzle: motion, no matter how well matched to the moving group's space velocity, is no guarantee the object is actually a young star of the appropriate age. Spectroscopic evidence of youth in the form of measurable lithium or low surface gravity features (weak lines of neutral potassium, sodium, and calcium, \citealt{Schlieder2012b}) are important independent measurements of youth apart from the motion assessment.

Using the South African Large Telescope (SALT), we set out to obtain evidence for stellar youth from radial velocities and spectroscopic parameters for an additional sample of stars from the TINYMO survey. In Section \ref{sec:sample}, we describe the sample selection process. In Section \ref{sec:observations} we detail the observational setup of the Red Side Spectrograph, our observational campaign, and our data reduction procedures. In Section \ref{sec:analysis}, we describe the spectroscopic measurements used to determine the radial velocities and gravity- and activity-sensitive spectral line measurements examined in Section \ref{sec:results}, and we discuss the individual young stars in more detail in Section \ref{sec:discussion}.

\section{Sample}
\label{sec:sample}

The sample of stars were drawn from the TINYMO survey \citep{Riedel2012}. In that survey, nearby low-proper-motion M dwarfs in the southern hemisphere were identified in the SuperCOSMOS Science Archive \citep{Hambly2001a}. To be sensitive to stars with proper motions less than 0.18\arcsec yr$^{-1}$ all the way down to objects with zero proper motion, an SQL query extracted stars using an {\it upper} limit on their motion between the photographic plates. Photometric distance relations calibrated to the SuperCOSMOS plate photometry \citep{Hambly2004} were used to identify stars within 25 parsecs. 

The excess luminosity of giant stars makes them appear much closer than they actually are, so a color-color cut was constructed using $v$\footnote{The average of the SuperCOSMOS B$_J$ and second epoch R filter is used as a surrogate $V$ magnitude.}$-K$\footnote{Throughout this paper, $K$ is the 2MASS $K_s$ filter} vs. $J-K$ color to filter out giants. This exploits a property of the $J-K$ color, in that when it is plotted against $V-K$ (or any other optical band - $K$), the locus of giant stars separates from main sequence dwarfs over the approximate spectral type range M1-M6 \citep{Riedel2012}. Consequently, most of the stars in the survey and this paper are of those spectral types, and most of the targets (though not all) have SuperCOSMOS proper motions less than 0.18\arcsec yr$^{-1}$. Further searches of catalogs like the General Catalog of Variable Stars \citep{Samus2010} and Catalog of Galactic Carbon Stars \citep{Alksnis2001}, low-resolution red optical spectroscopy collected at the Lowell Observatory 1.8m and CTIO 1.5m, and quality cuts on the photometric distance estimates reduced the sample of stars potentially within 25 parsecs to 651 objects.

As noted in \citet{Riedel2012}, these selection criteria biased the sample toward detecting young stars for two major reasons. First, the space velocities of the nearby young moving groups are clustered around the local standard of rest and thus their members fall within the proper motion selection criteria. Second, young M dwarfs are overluminous, which means that the photometric distance relations identify them as being closer than they really are, and they preferentially scatter into the 25 parsec sample.

In \citet{Riedel2012} and the subsequent work outlined in \citet{Riedel2014}, trigonometric parallaxes and spectroscopy were obtained for a subset of identified low-proper-motion M dwarfs whose photometric distances were estimated to place them within 15 parsecs of the Sun, with preference to stars exhibiting X-ray emission (from the ROSAT All-Sky Survey) as a sign of coronal activity.

In this paper, we present red optical spectroscopic observations of 79 stars from the TINYMO survey selection of 651 potential nearby M dwarfs regardless of activity level. The targets are bright M dwarfs published in \citet{Riedel2014} without radial velocities (as of 2013 when observations took place; several radial velocities were published afterwards), and TINYMO M dwarfs that have not been studied before. Targets were prioritized by their SuperCOSMOS plate $I_{59}$ magnitudes. They cover a photographic plate magnitude range of $I_{59}$=9.12 (NLTT 47004AB) to $I_{59}$=10.85 (SCR~1316-0858). As these are the brightest targets, this sample is expected to be biased toward closer stars, binaries, and extremely young stars. Together with \citet{Riedel2014}, 100 stars from the TINYMO sample have been followed up.

We have chosen to replace the astrometry and photometry from the \citet{Hambly2001a} SuperCOSMOS catalog with more recent and precise CCD-based measurements. SuperCOSMOS's astrometry is relative, not absolute, and contains proper motions forced to zero average on a per-photographic-field basis, which we have replaced with absolute ICRS-grid positions and proper motions from UCAC4 \citep{Zacharias2013} or PPMXL \citep{Roeser2010} where UCAC4 motions were not available. SuperCOSMOS's photometry, though internally consistent, is photographic plate measurements in blue, red, and infrared \citep{Hambly2001b}; we have instead elected to use Johnson $V$ CCD photometry from the AAVSO Photometric All Sky Survey (APASS) Data Release 9 \citep{Henden2016}, and $K$ near-infrared photometry from the Two Micron All Sky Survey (2MASS, \citealt{Cutri2003}). The adopted literature data are collected in Table \ref{tab:supplementary}.

\begin{deluxetable*}{lllcccrlrlrcrl}
\tabletypesize{\small}
\setlength{\tabcolsep}{0.04in}
\tablecolumns{12}
\tablecaption{Previously Published Astrometry and Photometry\label{tab:supplementary}}
\tablehead{ 
\colhead{Name}   & 
\colhead{RA}     &
\colhead{DEC}    &
\colhead{pos.}  &
\colhead{$\mu_{RA}$\tablenotemark{a}} & 
\colhead{$\mu_{DEC}$} &
\colhead{$\mu$} &
\multicolumn{2}{c}{$\pi$} &
\multicolumn{2}{c}{$V$}   &
\multicolumn{2}{c}{$K$} \\
\colhead{}       &
\multicolumn{2}{c}{(J2000 E2000)} &
\colhead{ref.}     &
\multicolumn{2}{c}{(\arcsec$yr^{-1}$)} &
\colhead{ref.}   &
\colhead{(mas)}  &
\colhead{ref.}   &
\colhead{(mag.)} &
\colhead{ref.}   & 
\colhead{(mag.)} &
\colhead{ref.}   }
\startdata
\object[2MASS J00172353-6645124]{SCR~0017-6645}   & 004.348112 & -66.753424 & 1 &  102.9$\pm$ 1.0 &  -15.0$\pm$ 1.0 & 1 & 25.61$\pm$1.73 & 3 & 12.49$\pm$0.04 & 7 & 7.70$\pm$0.02 & 1 \\
\object[GJ 2006 A]{GJ~2006A}        & 006.959305 & -32.551783 & 1 &   99.2$\pm$ 1.3 &  -61.3$\pm$ 2.6 & 1 & 30.97$\pm$1.76 & 3 & 12.82$\pm$0.06 & 7 & 8.01$\pm$0.03 & 1 \\
\object[GJ 2006 B]{GJ~2006B}        & 006.959810 & -32.556723 & 1 &  117.2$\pm$ 4.1 &  -31.5$\pm$ 5.8 & 1 & 30.97$\pm$1.76 & 3 & 13.14$\pm$0.04 & 7 & 8.12$\pm$0.03 & 1 \\
\object[HIP 3556]{HIP~3556}        & 011.367317 & -51.626090 & 1 &  100.3$\pm$ 1.3 &  -57.1$\pm$ 0.9 & 1 & 24.78$\pm$2.65 & 4 & 11.97$\pm$0.04 & 7 & 7.62$\pm$0.03 & 1 \\
\object[2MASS J01062266-6346391]{SCR~0106-6346}   & 016.594516 & -63.777545 & 1 &  150.3$\pm$ 1.3 &   65.0$\pm$ 1.3 & 1 &                &   & 13.40$\pm$0.04 & 7 & 8.39$\pm$0.03 & 1 \\
\object[GJ 3096]{{[PS78]}~190}    & 020.683483 & -25.785484 & 1 &   50.3$\pm$ 1.1 &    6.1$\pm$ 1.5 & 1 &                &   & 13.01$\pm$0.05 & 7 & 8.28$\pm$0.03 & 1 \\
\object[Barta 161 12]{BAR~161-12}     & 023.808013 & -07.214303 & 1 &   93.0$\pm$ 1.7 &  -48.0$\pm$ 2.2 & 1 & 33.70$\pm$0.26 & 5 & 13.43$\pm$0.04 & 7 & 8.08$\pm$0.03 & 1 \\
\object[UCAC3 154-4068]{GIC~138}        & 023.985392 & -13.429697 & 1 &  119.5$\pm$ 2.5 &  -21.5$\pm$ 3.2 & 1 &                &   & 13.36$\pm$0.01 & 7 & 8.81$\pm$0.02 & 1 \\
\object[L 173-39]{L~173-39}       & 027.108869 & -56.978227 & 1 &  255.6$\pm$ 8.0 &  -35.0$\pm$ 8.0 & 1 &                &   & 11.72$\pm$0.02 & 7 & 7.32$\pm$0.02 & 1 \\
\object[2MASS J01490591-5411571]{SCR~0149-5411}   & 027.274579 & -54.199205 & 1 &  120.0$\pm$ 1.4 &  -18.0$\pm$ 1.4 & 1 &                &   & 13.15$\pm$0.02 & 7 & 8.85$\pm$0.02 & 1 \\
\object[UCAC3 61-3820]{SCR~0152-5950}   & 028.076259 & -59.837995 & 1 &  109.2$\pm$ 1.8 &  -25.7$\pm$ 1.8 & 1 &                &   & 12.49$\pm$0.08 & 7 & 8.14$\pm$0.03 & 1 \\
\object[2MASS J02125819-5851182]{SCR~0212-5851}   & 033.242464 & -58.855051 & 1 &   87.7$\pm$ 1.3 &  -15.9$\pm$ 1.3 & 1 &                &   & 12.92$\pm$0.03 & 7 & 8.44$\pm$0.02 & 1 \\
\object[2MASS J02133021-4654505]{SCR~0213-4654}   & 033.375897 & -46.914036 & 1 &   42.5$\pm$ 1.0 &    4.9$\pm$ 1.0 & 1 &                &   & 13.78$\pm$0.07 & 7 & 8.60$\pm$0.02 & 1 \\
\object[BPS CS 22175-0005]{SCR~0215-0929}   & 033.995595 & -09.486749 & 1 &   96.6$\pm$ 1.9 &  -46.5$\pm$ 2.6 & 1 &                &   & 12.21$\pm$0.05 & 7 & 7.55$\pm$0.02 & 1 \\
\object[UCAC4 159-002053]{SCR~0220-5823}   & 035.214147 & -58.394755 & 1 &   97.3$\pm$ 2.0 &  -13.0$\pm$ 2.0 & 1 &                &   & 13.92$\pm$0.01 & 7 & 8.83$\pm$0.02 & 1 \\
\object[UCAC4 149-002104]{SCR~0222-6022}   & 035.683964 & -60.379890 & 1 &  137.4$\pm$ 1.7 &  -13.8$\pm$ 1.7 & 1 &                &   & 13.33$\pm$0.05 & 7 & 8.10$\pm$0.03 & 1 \\
\object[RBS 332]{2MASS~0236-5203} & 039.215438 & -52.051011 & 1 &  102.2$\pm$ 0.8 &    1.2$\pm$ 0.8 & 1 &                &   & 12.05$\pm$0.09 & 7 & 7.50$\pm$0.03 & 1 \\
\object[LP 886-73]{LP~886-73}      & 039.823509 & -26.821910 & 1 &   98.7$\pm$ 2.9 &  -40.2$\pm$ 1.3 & 1 &                &   & 14.33$\pm$0.05 & 7 & 8.75$\pm$0.02 & 1 \\
\object[UCAC3 112-6119]{SCR~0248-3404}   & 042.219172 & -34.073538 & 1 &   90.2$\pm$ 1.4 &  -23.7$\pm$ 1.4 & 1 &                &   & 13.64$\pm$0.02 & 7 & 8.40$\pm$0.03 & 1 \\
\object[UCAC2 6482611]{SCR~0254-5746}   & 043.526282 & -57.776673 & 1 &  102.9$\pm$ 1.1 &    7.2$\pm$ 1.2 & 1 &                &   & 13.37$\pm$0.03 & 7 & 8.83$\pm$0.02 & 1 \\
\object[GSC 08057-00342]{2MASS~0254-5108A}& 043.638184 & -51.142059 & 1 &   92.0$\pm$ 1.2 &  -11.9$\pm$ 1.2 & 1 &                &   & 12.07$\pm$0.03 & 7 & 7.79$\pm$0.03 & 1 \\
\object[2MASS J02564708-6343027]{SCR~0256-6343}   & 044.196132 & -63.717440 & 1 &   67.4$\pm$ 2.2 &    8.3$\pm$ 5.6 & 1 &                &   & 14.23          & 8 & 9.01$\pm$0.03 & 1 \\
\object[LP 831-35]{LP~831-35}      & 047.512712 & -23.691887 & 1 &   98.3$\pm$ 1.3 & -134.8$\pm$ 1.3 & 1 &                &   & 13.49$\pm$0.04 & 7 & 8.57$\pm$0.03 & 1 \\
\object[UCAC3 133-15653]{2MASS~0510-2340A}& 077.517787 & -23.678016 & 1 &   41.4$\pm$ 2.3 &  -13.3$\pm$ 1.1 & 1 &                &   & 13.04$\pm$0.03 & 7 & 8.36$\pm$0.02 & 1 \\
\object[RBS 626]{2MASS~0510-2340B}& 077.520402 & -23.670874 & 1 &   34.8$\pm$ 2.7 &  -13.8$\pm$ 1.5 & 1 &                &   & 13.29$\pm$0.05 & 7 & 8.54$\pm$0.02 & 1 \\
\object[1RXS J052241.4-060623]{SCR~0522-0606}   & 080.669559 & -06.106641 & 1 &   17.0$\pm$ 3.2 &  -21.1$\pm$ 3.3 & 1 &                &   & 14.27$\pm$0.06 & 7 & 9.13$\pm$0.02 & 1 \\
\object[2MASS J07115917-3510157]{SCR~0711-3510AB} & 107.996535 & -35.171050 & 1 &  -27.7$\pm$ 1.2 &  -57.9$\pm$ 1.6 & 1 &                &   & 13.65$\pm$0.06 & 7 & 8.79$\pm$0.02 & 1 \\
\object[UCAC4 417-046644]{SCR~0844-0637}   & 131.231937 & -06.623875 & 2 &  -58.3$\pm$ 5.3 & -126.4$\pm$ 5.3 & 2 &                &   & 13.38$\pm$0.05 & 7 & 8.51$\pm$0.02 & 1 \\
\object[LP 728-71]{LP~728-71}      & 148.174075 & -15.603822 & 1 & -117.0$\pm$ 1.2 & -135.2$\pm$ 1.3 & 1 &                &   & 13.46$\pm$0.05 & 7 & 8.51$\pm$0.02 & 1 \\
\object[UCAC3 118-135558]{SCR~1012-3124AB} & 153.037878 & -31.412579 & 1 &  -74.8$\pm$ 1.1 &   -9.4$\pm$ 1.0 & 1 & 18.54$\pm$1.74 & 3 & 13.42$\pm$0.07 & 7 & 7.99$\pm$0.03 & 1 \\
\object[TWA 3]{TWA~3ABCD}       & 167.616225 & -37.531102 & 1 & -105.9$\pm$ 0.9 &  -17.3$\pm$ 1.0 & 1 &                &   & 12.05$\pm$0.01 & 7 & 6.77$\pm$0.02 & 1 \\
\object[V* V1217 Cen]{SCR~1121-3845}   & 170.272849 & -38.754586 & 1 &  -66.6$\pm$ 1.5 &  -11.7$\pm$ 1.5 & 1 & 15.59$\pm$0.70 & 6 & 12.59$\pm$0.06 & 7 & 8.05$\pm$0.03 & 1 \\
\object[2MASS J11315526-3436272]{TWA~5ABC}        & 172.980251 & -34.607570 & 1 &  -79.6$\pm$ 0.8 &  -22.6$\pm$ 0.9 & 1 & 19.97$\pm$0.70 & 6 & 11.45$\pm$0.13 & 7 & 6.75$\pm$0.02 & 1 \\
\object[TWA 30]{RX~1132-3019}    & 173.076313 & -30.331070 & 1 &  -87.8$\pm$ 1.3 &  -25.2$\pm$ 1.3 & 1 &                &   & 14.41          & 8 & 8.77$\pm$0.02 & 1 \\
\object[CD-26 8623]{RX~1132-2651A}   & 173.171855 & -26.865554 & 2 &  -99.2$\pm$ 6.2 &  -32.2$\pm$ 6.2 & 2 & 21.28$\pm$1.01 & 3 & 12.27$\pm$0.02 & 7 & 7.43$\pm$0.02 & 1 \\
\object[2MASS J11453539-4055585]{SIPS~1145-4055}  & 176.398474 & -40.932576 & 1 & -277.2$\pm$ 8.0 & -131.9$\pm$ 8.0 & 1 &                &   & 14.22$\pm$0.04 & 1 & 8.79$\pm$0.02 & 1 \\
\object[LP 851-410]{LP~851-410}      & 179.582406 & -22.683258 & 1 & -105.1$\pm$ 1.9 &  -65.1$\pm$ 1.0 & 1 &                &   & 13.20$\pm$0.05 & 7 & 8.43$\pm$0.02 & 1 \\
\object[UCAC3 145-136405]{SCR~1200-1731}   & 180.006620 & -17.525233 & 1 &  -81.0$\pm$ 1.1 &  -24.6$\pm$ 1.2 & 1 &                &   & 13.83$\pm$0.06 & 7 & 8.47$\pm$0.03 & 1 \\
\object[2MASS J12072738-3247002]{2MASS~1207-3247} & 181.864072 & -32.783402 & 1 &  -70.4$\pm$ 1.4 &  -29.7$\pm$ 1.1 & 1 & 18.55$\pm$0.48 & 6 & 12.64$\pm$0.04 & 7 & 7.75$\pm$0.03 & 1 \\
\object[L 758-107]{L~758-107}       & 182.820600 & -19.972708 & 1 & -204.5$\pm$ 2.1 & -190.6$\pm$ 3.0 & 1 &                &   & 12.62$\pm$0.02 & 7 & 7.74$\pm$0.02 & 1 \\
\object[UCAC3 114-146748]{SCR~1230-3300}   & 187.720947 & -33.014119 & 1 & -156.7$\pm$ 0.8 &    0.0$\pm$ 0.8 & 1 &                &   & 12.56$\pm$0.03 & 7 & 8.09$\pm$0.02 & 1 \\
\object[2MASS J12333140-3641407]{SCR~1233-3641}   & 188.380858 & -36.694691 & 1 &  -55.8$\pm$ 0.9 &  -49.9$\pm$ 0.9 & 1 &                &   & 13.44$\pm$0.03 & 7 & 8.74$\pm$0.02 & 1 \\
SCR~1237-4021   & 189.301605 & -40.363386 & 1 &  -63.7$\pm$ 1.1 &  -29.1$\pm$ 1.1 & 1 &                &   & 13.50$\pm$0.06 & 7 & 8.52$\pm$0.02 & 1 \\
\object[UCAC4 315-070111]{SCR~1238-2703}   & 189.654646 & -27.059737 & 2 & -185.1$\pm$ 5.1 & -185.2$\pm$ 5.1 & 2 &                &   & 12.44$\pm$0.03 & 7 & 7.84$\pm$0.03 & 1 \\
\object[UCAC3 163-126512]{SCR~1316-0858}   & 199.168930 & -08.973762 & 1 &  -57.5$\pm$ 4.2 &  -56.2$\pm$ 7.7 & 1 &                &   & 14.57$\pm$0.06 & 7 & 9.20$\pm$0.02 & 1 \\
\object[UCAC4 396-055485]{SCR~1321-1052}   & 200.484680 & -10.869421 & 1 &  -66.6$\pm$ 3.1 &  -50.5$\pm$ 3.8 & 1 &                &   & 13.90$\pm$0.03 & 7 & 8.62$\pm$0.02 & 1 \\
\object[UCAC4 404-057550]{SCR~1421-0916}   & 215.359374 & -09.282748 & 1 & -135.3$\pm$ 3.3 &  -18.5$\pm$ 1.2 & 1 &                &   & 13.73$\pm$0.07 & 7 & 8.94$\pm$0.02 & 1 \\
\object[UCAC4 411-057663]{SCR~1421-0755}   & 215.391939 & -07.921291 & 1 &  -95.1$\pm$ 2.2 &  -86.4$\pm$ 1.2 & 1 &                &   & 13.61$\pm$0.01 & 7 & 8.63$\pm$0.02 & 1 \\
\object[UCAC3 98-153320]{SCR~1425-4113AB} & 216.371348 & -41.225645 & 1 &  -46.8$\pm$ 2.1 &  -49.2$\pm$ 1.7 & 1 & 14.94$\pm$0.96 & 3 & 12.62$\pm$0.05 & 7 & 7.61$\pm$0.02 & 1 \\
\object[CD-39 9080]{SCR~1438-3941}   & 219.651047 & -39.685075 & 1 & -109.8$\pm$ 1.0 & -104.5$\pm$ 2.2 & 1 &                &   & 12.73$\pm$0.01 & 7 & 8.52$\pm$0.02 & 1 \\
\object[LP 914-6]{LP~914-6}      & 220.092260 & -27.878378 & 1 & -127.9$\pm$ 8.0 & -202.2$\pm$ 8.0 & 1 &                &   & 13.62$\pm$0.01 & 7 & 8.75$\pm$0.02 & 1 \\
SCR~1521-2514   & 230.461609 & -25.236576 & 1 &  -32.4$\pm$ 1.4 &  -56.2$\pm$ 2.0 & 1 &                &   & 13.38$\pm$0.03 & 7 & 8.65$\pm$0.02 & 1 \\
\object[2MASS J17080882-6936186]{SCR~1708-6936}   & 257.036730 & -69.605169 & 1 &  -54.6$\pm$ 1.7 &  -81.1$\pm$ 1.7 & 1 &                &   & 13.16$\pm$0.02 & 7 & 8.20$\pm$0.02 & 1 \\
\object[UCAC2 4014059]{SCR~1816-6305}   & 274.211919 & -63.088775 & 1 & -117.5$\pm$ 6.1 &  -47.3$\pm$ 2.0 & 1 &                &   & 12.74$\pm$0.01 & 7 & 8.39$\pm$0.03 & 1 \\
\object[2MASS J18420694-5554254]{SCR~1842-5554A}  & 280.528984 & -55.907110 & 1 &    9.7$\pm$12.1 &  -81.2$\pm$ 2.8 & 1 &                &   & 13.59$\pm$0.14 & 7 & 8.58$\pm$0.02 & 1 \\
\object[UCAC2 11886946]{NLTT~47004AB}    & 282.172344 & -46.785495 & 2 &  196.6$\pm$ 3.2 &  125.6$\pm$ 3.2 & 2 &                &   & 11.60$\pm$0.02 & 7 & 6.99$\pm$0.04 & 1 \\
\object[UCAC3 42-239515]{SCR~1856-6922}   & 284.018227 & -69.366773 & 1 &  -10.8$\pm$ 1.4 & -115.6$\pm$ 2.3 & 1 &                &   & 12.44$\pm$0.04 & 7 & 7.70$\pm$0.02 & 1 \\
\object[UCAC4 178-221391]{WT~625}         & 286.334771 & -54.578257 & 1 &   86.2$\pm$ 8.0 & -213.3$\pm$ 8.0 & 1 &                &   & 13.34$\pm$0.01 & 1 & 8.55$\pm$0.02 & 1 \\
\object[2MASS J19225071-6310581]{SCR~1922-6310}   & 290.711309 & -63.182795 & 1 &   -7.9$\pm$16.7 &  -77.5$\pm$ 1.9 & 1 &                &   & 13.31$\pm$0.06 & 7 & 8.58$\pm$0.02 & 1 \\
\object[2MASS J19243494-3442392]{RX~1924-3442}    & 291.145631 & -34.710924 & 1 &   22.1$\pm$ 1.8 &  -71.7$\pm$ 1.8 & 1 &                &   & 14.24$\pm$0.06 & 7 & 8.79$\pm$0.03 & 1 \\
SCR~1926-5331   & 291.503116 & -53.524166 & 1 &   34.1$\pm$ 2.1 &  -87.4$\pm$ 2.1 & 1 &                &   & 14.03$\pm$0.13 & 7 & 8.68$\pm$0.02 & 1 \\
\object[UCAC3 132-425379]{SCR~1938-2416}   & 294.653739 & -24.282940 & 1 &   33.5$\pm$ 1.8 &   67.6$\pm$ 5.8 & 1 &                &   & 13.10$\pm$0.03 & 7 & 8.51$\pm$0.02 & 1 \\
\object[UCAC4 248-184370]{SCR~1951-4025}   & 297.899813 & -40.422480 & 2 &   40.8$\pm$14.0 & -186.1$\pm$14.0 & 2 &                &   & 13.55$\pm$0.02 & 7 & 8.71$\pm$0.02 & 1 \\
\object[UCAC2 2497991]{SCR~2004-6725A}  & 301.038323 & -67.419721 & 1 &    7.0$\pm$ 1.3 &  -84.5$\pm$ 2.4 & 1 &                &   & 13.11$\pm$0.02 & 7 & 8.48$\pm$0.02 & 1 \\
\object[2MASS J20042845-3356105]{2MASS~2004-3356} & 301.118608 & -33.936334 & 1 &   69.4$\pm$ 2.5 & -103.3$\pm$ 2.6 & 1 &                &   & 14.61$\pm$0.05 & 7 & 9.17$\pm$0.02 & 1 \\
\object[2MASS J20085368-3519486]{SCR~2008-3519}   & 302.223697 & -35.330161 & 1 &   49.4$\pm$ 1.3 &  -76.7$\pm$ 1.3 & 1 &                &   & 13.52$\pm$0.06 & 7 & 8.32$\pm$0.03 & 1 \\
\object[UCAC3 124-580676]{SCR~2010-2801AB} & 302.500154 & -28.028066 & 1 &   40.7$\pm$ 3.0 &  -62.0$\pm$ 1.7 & 1 & 20.85$\pm$1.33 & 3 & 12.98$\pm$0.02 & 7 & 7.73$\pm$0.03 & 1 \\
\object[L 755-19]{L~755-19}       & 307.181835 & -11.475196 & 2 &  166.4$\pm$ 5.2 &  -93.3$\pm$ 5.2 & 2 & 53.18$\pm$1.67 & 3 & 12.52$\pm$0.04 & 7 & 7.50$\pm$0.03 & 1 \\
\object[UCAC3 39-193865]{SCR~2107-7056}   & 316.843702 & -70.936854 & 1 &   27.5$\pm$ 1.5 &  -91.6$\pm$ 1.4 & 1 &                &   & 13.86$\pm$0.08 & 7 & 8.90$\pm$0.02 & 1 \\
\object[UCAC4 385-156333]{SCR~2107-1304}   & 316.903307 & -13.082831 & 1 &   59.4$\pm$ 1.3 &  -86.0$\pm$ 3.0 & 1 &                &   & 12.64$\pm$0.08 & 7 & 7.84$\pm$0.03 & 1 \\
\object[UCAC4 135-176640]{LEHPM~1-4147}    & 323.717454 & -63.018961 & 2 &  105.9$\pm$14.0 & -134.9$\pm$14.0 & 2 &                &   & 13.47$\pm$0.04 & 7 & 8.74$\pm$0.03 & 1 \\
\object[CRTS J220439.8-071133]{SCR~2204-0711}   & 331.165971 & -07.192684 & 1 &   -3.1$\pm$ 1.3 &   -4.2$\pm$ 1.4 & 1 &                &   & 12.27$\pm$0.21 & 7 &-8.49$\pm$0.02 & 1 \\
\object[RBS 1877]{SCR~2237-2622}   & 339.312287 & -26.375884 & 1 &  145.4$\pm$ 1.5 &  -11.8$\pm$ 1.5 & 1 &                &   & 13.33$\pm$0.04 & 7 & 8.31$\pm$0.02 & 1 \\
\object[UCAC4 395-130914]{SIPS~2258-1104}  & 344.568493 & -11.071400 & 1 &  106.8$\pm$ 2.9 &   -5.9$\pm$ 2.9 & 1 &                &   & 12.97$\pm$0.05 & 7 & 8.24$\pm$0.03 & 1 \\
\object[UCAC4 184-216216]{LEHPM~1-5404}    & 345.456139 & -53.285470 & 2 &  151.1$\pm$16.1 & -234.0$\pm$16.1 & 2 &                &   & 13.11$\pm$0.04 & 7 & 8.50$\pm$0.02 & 1 \\
\object[UCAC4 110-129613]{SCR~2328-6802}   & 352.240163 & -68.042788 & 1 &   66.8$\pm$ 1.9 &  -67.1$\pm$ 1.7 & 1 &                &   & 13.02$\pm$0.04 & 7 & 8.38$\pm$0.02 & 1 \\
\object[LTT 9582]{LTT~9582}       & 353.000790 & -39.293587 & 2 &  193.4$\pm$17.9 & -178.4$\pm$17.9 & 2 &                &   & 12.96$\pm$0.07 & 7 & 8.02$\pm$0.02 & 1 \\
\object[G 275-71]{G~275-71}       & 353.953102 & -24.319238 & 1 &  106.5$\pm$ 1.1 & -120.7$\pm$ 1.0 & 1 &                &   & 13.71$\pm$0.02 & 7 & 8.77$\pm$0.02 & 1 \\
\object[UCAC4 249-184943]{LEHPM~1-6053}    & 355.099384 & -40.363084 & 2 &  262.5$\pm$13.3 & -131.0$\pm$13.3 & 2 &                &   & 12.94$\pm$0.02 & 7 & 8.34$\pm$0.03 & 1 \\
\enddata
\tablecomments{The kinematic data used by LACEwING (in section \ref{sec:results}) and photometric data used in our youth analysis. Literature RVs were not used even 
where they exist. 1.) UCAC4 (includes APASS DR6 $g'r'i'BV$ and 2MASS $JHK_s$) \citet{Zacharias2013}, 2.) PPMXL \citet{Roeser2010}, 
3.) \citet{Riedel2014}, 4.) \citet{van-Leeuwen2007}, 5.) \citet{Shkolnik2012}, 6.) \citet{Weinberger2013}, 7.) \citet{Henden2016} 8.) \citet{Zacharias2005}. {\it The online machine-readable version of this table is a merged table including Table \ref{tab:RVtable}, Table \ref{tab:youth}, and Table \ref{tab:sample}.}}
\tablenotetext{a}{$\mu_{RA}$ is actually $\mu_{RA \cos{DEC}}$ everywhere it appears.}
\end{deluxetable*}

\section{Observations}
\label{sec:observations}
The South African Large Telescope (SALT) is an 11-meter telescope hosted at the South African Astronomical Observatory (SAAO) in Sutherland, South Africa. It is based on the design of the McDonald Observatory Hobby-Eberly Telescope \citep{Buckley2006} and shares the segmented mirror and fixed-altitude design of that telescope, yielding an effective mirror diameter of 9.2 meters. The large collecting area and advantageous position in the southern hemisphere (where most of the NYMGs are concentrated) make SALT an efficient and effective means of collecting high SNR low resolution spectra of young M dwarfs even under bad weather conditions. It was thus possible to schedule usable observations at all times despite SALT's fixed altitude and the associated maximum tracking time requirements for objects at mid-southern declinations.

We have obtained low resolution optical spectroscopy from the SALT telescope and the Robert Stoble Spectrograph (RSS, \citealt{Kobulnicky2003}), which provides optical spectroscopy between 3200-9000\AA~with a resolving power of up to 6000, depending on slit width.

Observations were conducted in semesters 2013A and 2013B, utilizing bad-weather time; on many occasions only a single star was observed each night. At all times, the PG1800 filter was used with a 1\arcsec~slit and the PC04600 grating at an angle of 40.25 degrees. In this mode, RSS delivers spectra covering 6500-7800\AA~with a resolving power of R=5000, covering the H$\alpha$ 6563\AA~line, the Lithium 6708\AA~doublet, and the Potassium 7699\AA~line. The RSS chip is split into 3 equal segments, which can be reduced separately; in this way, the chips cover roughly 6500-6900\AA, 6950-7350\AA, and 7400-7800\AA. During the 2013A semester, a Neon arc lamp was used; in 2013B, this was changed to Xenon to obtain more evenly spaced lines.

Two spectra were taken per visit in a 2x2 binning mode with slow readout and with exposure times calculated for 5\arcsec~seeing conditions (in practice, the seeing was never that bad). With these conditions and bright targets (exposures were never longer than 620 seconds), signal to noise ratios of over 100 were generally reached.

In total, there are 165 spectra of the 79 stars: SCR~2237-2622 was only observed once; two stars (SCR~1816-6305, 2MASS~2004-3356) were observed three times, three stars (2MASS~0510-2340B, 2MASS~1207-3247, SCR~1842-5554A) were observed four times, and the remainder were observed twice. Twelve spectra of telluric standard stars were also taken. The entire observing program totaled 140,000 seconds (39 hours) of observing time.

Data were reduced with PyRAF and the Astropy astronomical python package \citep{Astropy2013}, using standard IRAF long-slit reduction techniques. The individual RSS chips were treated separately, to increase the accuracy of radial velocity calibrations. The wavelength calibration error obtained by IRAF was around 0.04\AA~for both Xenon and Neon arc lamps. Of all the flux standards observed, only Hiltner 600, observed in 2013B, was taken with sufficient signal to noise to be used in reductions. 

\section{Analysis}
\label{sec:analysis}

\begin{figure}
\centering
\includegraphics[width=0.5\textwidth]{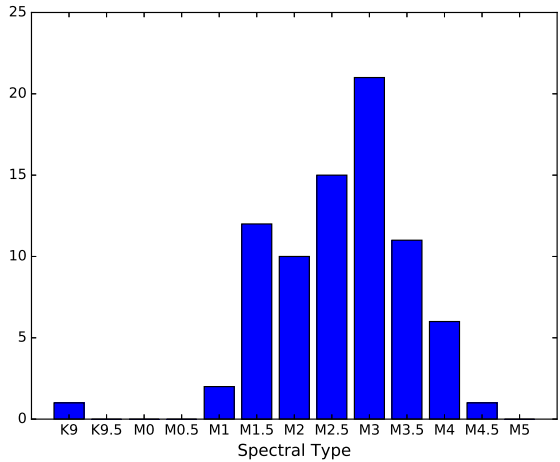}
\label{fig:sptypehist}
\caption{Distribution of spectral types of objects in the sample.}
\end{figure}

Spectral types (Table \ref{tab:RVtable}) were determined using the MATCHSTAR code \citep{Riedel2014}, an automated template matching code which compares red optical spectra to RECONS K and M standard star spectra  \citep{Kirkpatrick1991,Henry1994,Henry2002} between 6000\AA-9000\AA, after trimming and degrading resolution so that input and standard spectra overlap. With the prioritization of the brightest M dwarfs in the sample, it is unsurprising that most of the objects skew toward the hotter M dwarfs - M1.0V-M3.0V. The hottest star is SCR~2204-0711, K9.0Ve (though see Section \ref{sec:giants}); the coolest star is RX~1132-3019, M4.5Ve (Figure \ref{fig:sptypehist}). Spectral types are given in Table \ref{tab:RVtable}.

\subsection{Radial velocities}
\label{sec:RVs}

Radial velocities (RVs) were measured from the SALT data using the same code developed for and used in \citet{Faherty2016}. The code cross-correlates the spectrum of a star of unknown RV, to one of known RV and the same spectral type. The process is repeated 1000 times, adding random Gaussian noise scaled to the per-pixel flux uncertainties of both the known and unknown star's spectrum, in order to quantify the effect of noise on the radial velocity measurement. The RV results of the 1000 iterations are then binned into a histogram and fit with a Gaussian to determine the mean radial velocity and uncertainty.

As noted in \citet{Faherty2016}, this cross-correlation technique still under-estimates the true uncertainties in the measured RV. To accommodate systematic errors in the data, we cross-correlate the spectra against multiple comparison stars and combine the results with a weighted standard deviation. Ideally, the comparison stars would be RV standards, but we did not observe any radial velocity standards. Instead, we selected thirteen stars with previously measured RVs with uncertainties less than 2 km s$^{-1}$ for use as velocity comparisons. We measured the RVs of the comparison stars relative to each other other (the stars marked ``C'' in Table \ref{tab:RVtable}) to demonstrate the accuracy and precision of the radial velocities.

\begin{figure*}
\centering
\includegraphics[width=\textwidth]{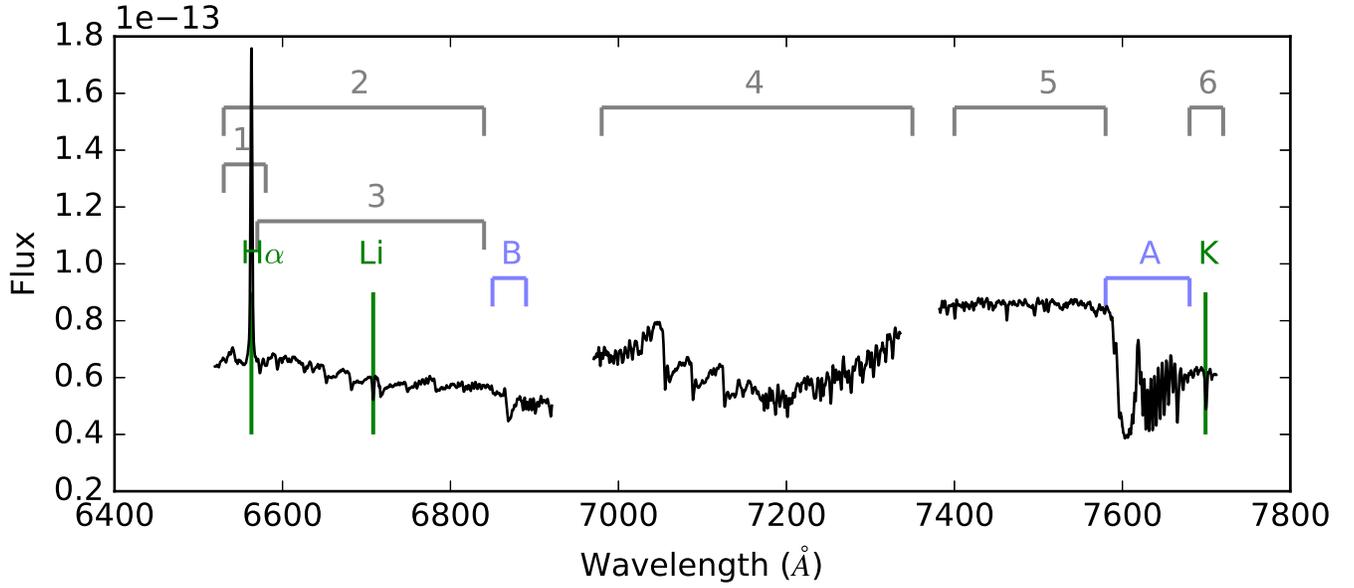}
\label{fig:regions}
\caption{Spectrum of RXJ~1132-2651A (TWA 8A) showing the spectral regions considered for RV fits, the spectral lines measured, and the regions used for the telluric correction to the atmospheric A and B bands.}
\end{figure*}

Six spectral regions were considered for RV measurement (Figure \ref{fig:regions}). (1) The first chip, in a 50\AA~region surrounding the H$\alpha$ line (6530-6580\AA), (2) the first chip, truncated at the atmospheric B band (6500-6840\AA), (3) the first chip, with both the H$\alpha$ and atmospheric B band removed (6570-6840\AA), (4) the second chip, covering 6980-7350\AA, (5) the third chip, blueward of the atmospheric A band (7400-7580\AA), and (6), the third chip, redward of the atmospheric A band. (7680-7720\AA). Initial RV measurements showed the original wavelength calibrations to both the neon and xenon lamps were insufficient for RV work. RVs of our comparison stars were typically discrepant from published values by over 10 km s$^{-1}$, even after being combined in weighted standard deviations. 

We investigated the possibility of improving our precision by using the atmospheric A and B bands (regions shown in Figure \ref{fig:regions}) to correct the wavelength solution. The first procedure attempted involved independently cross-correlating chip 1 and chip 3 to an atmospheric template spectrum \citep{Hinkle2003} to obtain zero-point corrections to the wavelength solution, before applying the heliocentric correction. This initial procedure produced a significant improvement in accuracy, but prevented the use of chip 2, which lacks prominent atmospheric features. A further improvement was made using the centers of the A and B bands to derive a linear correction as a function of input wavelength. This produced significantly better accuracy than the zero point correction, and it was found that the best results came from the second chip (region 4), where precisions of 3-6 km s$^{-1}$ have been achieved. All RV results in Table \ref{tab:RVtable} derive from spectral region 4 with the two-point linear wavelength correction.

While it is known that spectral type matches are important to obtain precise RVs, all of our targets have very similar spectral types, and our precision appears to be low enough that those spectral morphological differences do not affect our results. Instead, we found three stars -- SCR~0017-6645, SCR~0152-5950, and 2MASS~1207-3247 -- that produced uniformly low accuracy and low precision results in every cross-correlation. These were removed from consideration, and the radial velocity results are based on the weighted mean and weighted standard deviation of the other 10 stars with known RVs. The RV results and uncertainties in Table \ref{tab:RVtable} are thus the weighted standard deviation of between 10 and 40 measurements, depending on the number of spectra for the target. Of the 34 total stars in the sample with existing radial velocity measurements, 24 of our measurements (70\%) are within 1-$\sigma$ of those reported errors and 29 (85\%) are within 2-$\sigma$. Given the small number of RV crossmatches, we believe this demonstrates the accuracy of our RVs despite our relatively low precision.

\begin{deluxetable*}{llcccrrc}
\tabletypesize{\small}
\tablewidth{0pt}
\tablecolumns{8}
\tablecaption{Radial Velocities and Spectral Types\label{tab:RVtable}}

\tablehead{ 
\colhead{C\tablenotemark{a}}      &
\colhead{Name}   & 
\colhead{RA}     &
\colhead{DEC}    &
\colhead{Spectral} &
\colhead{R.V.}   &
\multicolumn{2}{c}{Literature R.V.} \\
\colhead{}                  &
\colhead{}                  &
\multicolumn{2}{c}{(J2000)} &
\colhead{Type}              &
\colhead{(km s$^{-1}$)}      &
\colhead{(km s$^{-1}$)}      &
\colhead{ref.}  }
\startdata
X & SCR~0017-6645    & 00 17 23.52 & -66 45 12.5 & M3.0 Ve  &   $-$4.8$\pm$5.6 & $+$11.4$\pm$0.8 & 1 \\
  & GJ~2006A         & 00 27 50.24 & -32 33 06.1 & M3.5 Ve  &  $-$24.9$\pm$4.6 &                 &   \\
  & GJ~2006B         & 00 27 50.36 & -32 33 23.9 & M3.5 Ve  &   $+$6.9$\pm$6.1 &                 &   \\
  & HIP~003556       & 00 45 28.15 & -51 37 34.0 & M1.5 Ve  &   $-$0.7$\pm$4.9 &  $-$1.6$\pm$20  & 1 \\
  & SCR~0106-6346    & 01 06 22.67 & -63 46 39.1 & M3.0 Ve  &  $+$13.3$\pm$5.2 &                 &   \\
  & {[PS78]}~190     & 01 22 44.04 & -25 47 07.8 & M3.0 Ve  &   $+$3.6$\pm$5.4 &                 &   \\
  & BAR~161-12      & 01 35 13.94 & -07 12 51.8 & M4.0 Ve  &  $+$19.0$\pm$5.8 & $+$11.7$\pm$5.3 & 2 \\
  & GIC~138         & 01 35 56.45 & -13 25 47.3 & M1.5 Ve  &  $-$26.4$\pm$7.2 &                 &   \\
  & L~173-39        & 01 48 26.17 & -56 58 41.5 & M1.5 Ve  &  $+$33.6$\pm$5.5 &                 &   \\
  & SCR~0149-5411    & 01 49 05.92 & -54 11 57.2 & M1.0 Ve  &   $+$2.1$\pm$4.1 &                 &   \\
X & SCR~0152-5950    & 01 52 18.31 & -59 50 16.8 & M2.0 Ve  &  $+$14.0$\pm$5.3 &  $+$7.9$\pm$1.6 & 1 \\
  & SCR~0212-5851    & 02 12 58.20 & -58 51 18.2 & M2.0 Ve  &   $+$4.5$\pm$5.0 &  $+$9.1$\pm$0.8 & 3 \\
  & SCR~0213-4654    & 02 13 30.22 & -46 54 50.5 & M3.0 Ve  &   $+$9.4$\pm$6.0 & $+$14.3$\pm$2.0 & 4 \\
  & SCR~0215-0929    & 02 15 58.93 & -09 29 12.2 & M2.5 Ve  &  $+$15.4$\pm$5.5 & $+$10.1$\pm$0.6 & 3 \\
  & SCR~0220-5823    & 02 20 51.39 & -58 23 41.1 & M3.5 Ve  &  $+$18.3$\pm$6.3 & $+$12.1$\pm$0.6 & 3 \\
  & SCR~0222-6022    & 02 22 44.17 & -60 22 47.6 & M3.5 Ve  &  $+$28.6$\pm$7.0 & $+$16.2$\pm$1.5 & 3 \\
C & 2MASS~0236-5203  & 02 36 51.71 & -52 03 03.7 & M2.0 Ve  &  $+$12.9$\pm$5.5 & $+$16.0$\pm$0.1 & 5 \\
  & LP~886-73       & 02 39 17.64 & -26 49 18.9 & M4.0 Ve  &  $-$16.5$\pm$6.4 & 		     &   \\
C & SCR~0248-3404    & 02 48 52.62 & -34 04 24.7 & M3.5 Ve  &  $+$14.5$\pm$5.5 & $+$14.6$\pm$0.3 & 4 \\
  & SCR~0254-5746    & 02 54 06.31 & -57 46 36.1 & M2.5 V   &   $+$0.1$\pm$5.0 & 		     &   \\
  & 2MASS~0254-5108A & 02 54 33.17 & -51 08 31.4 & M1.5 Ve  &  $+$13.0$\pm$5.1 & $+$13.8$\pm$0.4 & 3 \\
  & SCR~0256-6343    & 02 56 47.09 & -63 43 02.8 & M4.0 Ve  &  $-$10.9$\pm$6.0 & $+$16.2$\pm$3.4 & 4 \\
  & LP~831-35       & 03 10 03.07 & -23 41 31.0 & M3.5 Ve  &  $+$25.5$\pm$6.1 & 		     &   \\
  & 2MASS~0510-2340A & 05 10 04.27 & -23 40 40.7 & M3.0 Ve  &  $+$18.8$\pm$5.2 & $+$24.2$\pm$0.2 & 4 \\
  & 2MASS~0510-2340B & 05 10 04.88 & -23 40 14.9 & M2.5 Ve  &  $+$15.6$\pm$7.5 & $+$23.8$\pm$0.5 & 4 \\
  & SCR~0522-0606    & 05 22 40.70 & -06 06 23.9 & M2.5 Ve  &   $-$1.5$\pm$5.0 & 		     &   \\
  & SCR~0711-3510AB  & 07 11 59.17 & -35 10 15.7 & M3.0 Ve  &   $+$4.1$\pm$5.0 & 		     &   \\
  & SCR~0844-0637    & 08 44 55.66 & -06 37 26.0 & M2.0 Ve  &  $-$18.0$\pm$2.0 & 		     &   \\
  & LP~728-71       & 09 52 41.77 & -15 36 13.7 & M2.5 V   &   $-$5.6$\pm$3.4 & 		     &   \\
  & SCR~1012-3124AB  & 10 12 09.08 & -31 24 45.2 & M3.5 Ve  &  $+$14.6$\pm$5.5 &$+$14.69$\pm$0.53& 6 \\
C & TWA~3ABCD        & 11 10 27.88 & -37 31 52.0 & M3.5 Ve  &  $+$14.5$\pm$2.5 & $+$15.6$\pm$0.2 & 5 \\
C & SCR~1121-3845    & 11 21 05.49 & -38 45 16.4 & M1.0 Ve  &   $+$9.5$\pm$2.2 & $+$12.7$\pm$1.0 & 5 \\
  & TWA~5ABC         & 11 31 55.26 & -34 36 27.3 & M1.5 Ve  &  $+$10.2$\pm$2.2 & $+$12.7$\pm$3.8 & 5 \\
C & RX~1132-3019     & 11 32 18.31 & -30 19 51.8 & M4.5 Ve  &  $+$15.8$\pm$4.7 & $+$12.3$\pm$1.5 & 7 \\
C & RX~1132-2651A    & 11 32 41.25 & -26 51 55.9 & M2.0 Ve  &   $+$7.8$\pm$2.5 & $+$8.68$\pm$0.02& 1 \\
  & SIPS~1145-4055   & 11 45 35.70 & -40 55 57.0 & M2.5 Ve  &  $+$14.2$\pm$6.2 & 		     &   \\
  & LP~851-410       & 11 58 19.78 & -22 40 59.7 & M2.5 Ve  &  $-$36.7$\pm$4.3 & 		     &   \\
  & SCR~1200-1731    & 12 00 01.60 & -17 31 30.8 & M3.5 Ve  &  $+$20.1$\pm$2.4 & 		     &   \\
X & 2MASS~1207-3247  & 12 07 27.38 & -32 47 00.3 & M2.5 Ve  &  $-$16.4$\pm$2.9 &  $+$8.5$\pm$1.2 & 8 \\
C & L~758-107        & 12 11 16.95 & -19 58 21.7 & M2.5 V   &  $-$12.0$\pm$4.9 &$-$9.226$\pm$0.1 & 9 \\
  & SCR~1230-3300    & 12 30 53.02 & -33 00 50.8 & M1.5 Ve  &  $+$20.7$\pm$2.2 & 		     &   \\
  & SCR~1233-3641    & 12 33 31.40 & -36 41 40.8 & M2.0 Ve  &   $+$6.7$\pm$4.4 & 		     &   \\
  & SCR~1237-4021    & 12 37 12.38 & -40 21 48.1 & M2.5 Ve  &  $+$11.3$\pm$2.4 & 		     &   \\
  & SCR~1238-2703    & 12 38 37.13 & -27 03 35.0 & M1.5 Ve  &   $+$0.5$\pm$3.3 &  $+$9.9$\pm$0.2 & 4 \\
  & SCR~1316-0858    & 13 16 40.54 & -08 58 25.6 & M3.0 Ve  &   $+$6.1$\pm$4.9 & 		     &   \\
  & SCR~1321-1052    & 13 21 56.31 & -10 52 09.9 & M3.5 Ve  &  $+$26.5$\pm$5.6 &  $-$4.2$\pm$2.2 & 4 \\
  & SCR~1421-0916    & 14 21 26.24 & -09 16 58.0 & M2.0 V   &   $-$4.3$\pm$4.3 & 		     &   \\
  & SCR~1421-0755    & 14 21 34.06 & -07 55 16.6 & M3.0 V   &   $-$5.2$\pm$7.2 & 		     &   \\
  & SCR~1425-4113AB  & 14 25 29.20 & -41 13 32.0 & M3.0 Ve  &  $+$12.4$\pm$3.7 & 		     &   \\
  & SCR~1438-3941    & 14 38 36.40 & -39 41 04.0 & M1.5 V   &  $+$36.6$\pm$5.9 & 		     &   \\
  & LP~914-6         & 14 40 22.30 & -27 52 39.0 & M3.0 Ve  &  $+$17.5$\pm$4.1 & 		     &   \\
  & SCR~1521-2514    & 15 21 50.80 & -25 14 11.0 & M1.5 Ve  &   $-$0.8$\pm$2.9 & 		     &   \\
  & SCR~1708-6936    & 17 08 09.00 & -69 36 18.0 & M3.0 Ve  &  $+$11.3$\pm$3.4 &  $+$9.4$\pm$3.5 & 4 \\
  & SCR~1816-6305    & 18 16 51.10 & -63 05 19.0 & M1.5 V   &  $+$33.5$\pm$0.9 & 		     &   \\
  & SCR~1842-5554A   & 18 42 06.95 & -55 54 25.5 & M3.0 Ve  &   $+$9.8$\pm$4.9 &  $+$0.3$\pm$0.5 & 4 \\
  & NLTT~47004AB     & 18 48 41.10 & -46 47 10.0 & M2.5 Ve  &  $+$21.0$\pm$4.0 & 		     &   \\
  & SCR~1856-6922    & 18 56 04.40 & -69 21 59.0 & M3.0 V   &  $+$21.1$\pm$4.0 & 		     &   \\
  & WT~625           & 19 05 20.20 & -54 34 40.0 & M3.0 Ve  &  $-$15.6$\pm$4.3 & 		     &   \\
  & SCR~1922-6310    & 19 22 50.70 & -63 10 57.0 & M3.0 Ve  &   $+$5.7$\pm$2.1 &  $+$6.5$\pm$1.6 & 4 \\
C & RX~1924-3442     & 19 24 34.95 & -34 42 39.4 & M4.0 Ve  &   $-$7.7$\pm$5.7 &  $-$3.7$\pm$0.2 & 4 \\
  & SCR~1926-5331    & 19 26 00.75 & -53 31 26.9 & M4.0 Ve  &   $-$7.6$\pm$6.0 &  		     &   \\
  & SCR~1938-2416    & 19 38 36.90 & -24 17 00.0 & M2.0 Ve  &  $+$14.4$\pm$3.9 &  		     &   \\
  & SCR~1951-4025    & 19 51 35.90 & -40 25 18.0 & M1.5 Ve  &  $+$20.6$\pm$5.9 &  		     &   \\
  & SCR~2004-6725A   & 20 04 09.20 & -67 25 09.0 & M2.5 Ve  &  $+$10.4$\pm$3.5 &  		     &   \\
  & 2MASS~2004-3356  & 20 04 28.47 & -33 56 10.7 & M4.0 Ve  &  $-$16.1$\pm$3.8 & 		     &   \\
  & SCR~2008-3519    & 20 08 53.60 & -35 19 47.0 & M3.0 Ve  &   $-$5.9$\pm$6.3 & 		     &   \\
  & SCR~2010-2801AB  & 20 10 00.03 & -28 01 41.2 & M3.0 Ve  &   $+$1.0$\pm$4.2 &  $-$5.8$\pm$0.6 & 4 \\ 
  & L~755-19         & 20 28 43.40 & -11 28 29.0 & M1.5 Ve  &  $-$31.2$\pm$5.8 & 		     &   \\
  & SCR~2107-7056    & 21 07 22.48 & -70 56 12.6 & M3.0 Ve  &   $+$3.3$\pm$4.6 & 		     &   \\
  & SCR~2107-1304    & 21 07 36.70 & -13 04 56.0 & M3.0 Ve  &   $-$3.7$\pm$4.8 &  $-$2.3$\pm$0.5 & 4 \\
  & LEHPM~1-4147     & 21 34 52.00 & -63 01 06.0 & M2.5 V   &  $+$12.4$\pm$4.0 &  		     &   \\
  & SCR~2204-0711\tablenotemark{b}&22 04 39.83&-07 11 33.7& &$-$204.9$\pm$12.1 &                 &   \\
  & SCR~2237-2622    & 22 37 14.95 & -26 22 33.3 & M3.5 Ve  &   $+$3.4$\pm$5.0 & 		     &   \\
C & SIPS~2258-1104   & 22 58 16.44 & -11 04 17.1 & M3.0 Ve  &  $+$20.2$\pm$5.5 & $+$16.0$\pm$0.2 & 2 \\
  & LEHPM~1-5404     & 23 01 49.48 & -53 17 07.7 & M2.0 V   &  $+$12.3$\pm$6.7 & 		     &   \\
  & SCR~2328-6802    & 23 28 57.64 & -68 02 33.9 & M2.5 Ve  &   $+$3.5$\pm$5.1 &  $+$8.0$\pm$1.5 & 3 \\
C & LTT~9582         & 23 32 00.19 & -39 17 36.9 & M3.0 Ve  &   $+$8.7$\pm$5.3 & $+$11.6$\pm$0.7 & 1 \\
  & G~275-71         & 23 35 48.75 & -24 19 09.3 & M2.5 V   &   $-$2.9$\pm$7.7 & 		     &   \\
  & LEHPM~1-6053     & 23 40 23.85 & -40 21 47.1 & M2.0 Ve  &  $+$25.2$\pm$5.7 & 		     &   \\
\enddata
\tablecomments{Literature RV sources: 1.) \citet{Malo2013}, 2.) \citet{Shkolnik2012}, 3.) \citet{Kraus2014a}, 4.) \citet{Malo2014a}, 5.) \citet{Torres2006}, 6.) \citet{Riedel2014}, 7.) \citet{Looper2010}, 8.) \citet{Schneider2012}, 9.) \citet{Nidever2002}}
\tablenotetext{a}{Stars are noted as ``C'' if their spectra were used as comparisons for the RV fitting, and ``X'' if they were rejected as comparisons.}
\tablenotetext{b}{Giant, see \ref{sec:giants}}
\end{deluxetable*}

\subsection{Spectral Line Measurements}

To evaluate spectroscopic signatures of youth, we measure equivalent widths of the H$\alpha$ (6563 \AA) emission line, the unresolved Li (6708 \AA) doublet, and the K {\sc i} (7699 \AA) absorption line for all 177 low-resolution optical spectra taken with SALT. To aid us in these measurements, we employ PHEW: PytHon Equivalent Widths \citep{PHEW}, which is based on the Pythonic spectroscopic line analysis toolkit PySpecKit\footnote{\url{http://pyspeckit.bitbucket.org/html/sphinx/index.html}}  \citep{Ginsburg2011}.

\begin{figure}
\centering
\includegraphics[width=3.5in]{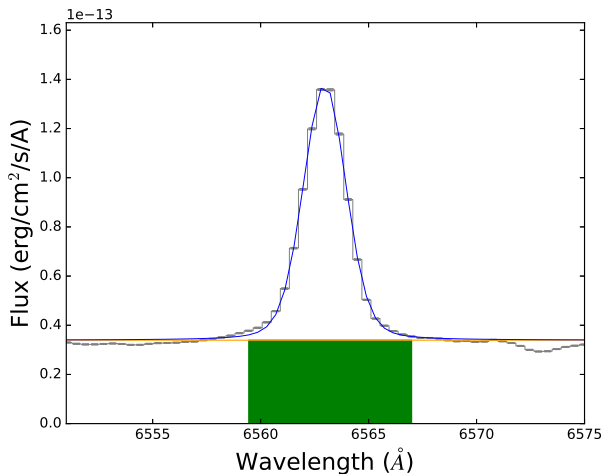}
\label{fig:RX1132_Ha}
\caption{H$\alpha$ (6563 \AA) EW of RX~1132-2651A (TWA 8A), as an example. The pseudocontinuum (yellow) and Voigt profile (blue) are fit to the observed spectrum (grey). The green rectangle approximates the EW.}
\end{figure}

\begin{figure}
\centering
\includegraphics[width=3.5in]{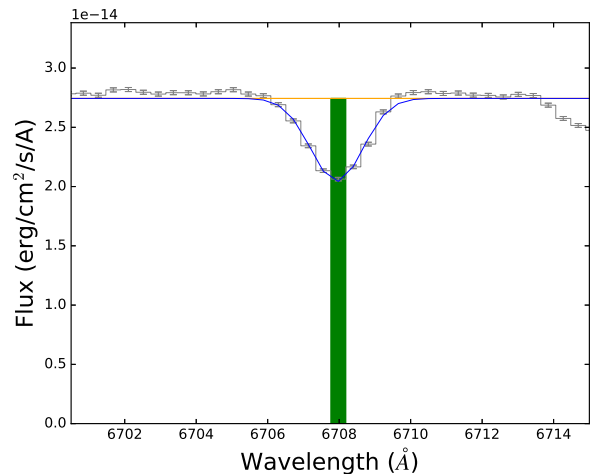}
\caption{Li (6708 \AA) EW of RX~1132-2651A (TWA 8A), as an example. The pseudocontinuum and Voigt profile are shown according to Figure \ref{fig:RX1132_Ha}. The green rectangle approximates the EW.}
\label{fig:RX1132_Li}
\end{figure}

\begin{figure}	
\centering
\includegraphics[width=3.5in]{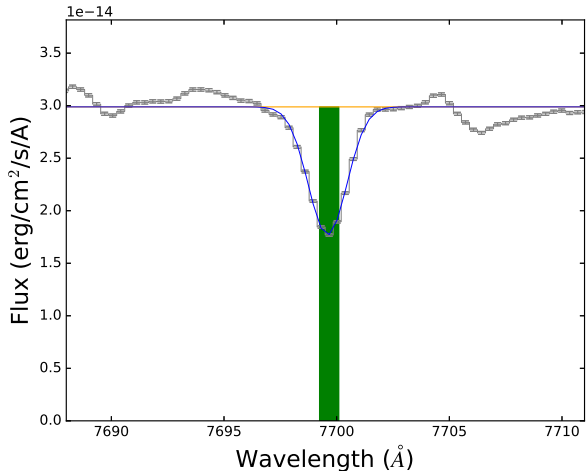}
\caption{K {\sc I} (7699 \AA) EW of RX~1132-2651A (TWA 8A), as an example. The EW itself is approximated by the green rectangle. A Voigt profile and pseudocontinuum fit are shown as in Figures \ref{fig:RX1132_Ha} and \ref{fig:RX1132_Li}.}
\label{fig:RX1132_KI}
\end{figure}

For each case, we fit a 0th-order baseline to the average flux of the nearby pseudocontinuum, and set the line window to be between 6550-6580 \AA~(Figure \ref{fig:RX1132_Ha}), 6700-6715 \AA~(Figure \ref{fig:RX1132_Li}), and 7685-7711 \AA~(Figure \ref{fig:RX1132_KI}) for the H$\alpha$, Li, and K {\sc i} lines, respectively. We then fit a Voigt profile to the spectral line in each window. (For features that do not have well-defined wings at this resolution, equivalent widths were measured with a Gaussian profile fit to the feature.) The equivalent width is then calculated by integrating the pseudocontinuum level minus the spectrum over the selected range. Uncertainties were estimated through Monte Carlo analysis of 500 iterations, for every spectrum. We have combined all measurements for a given star with a weighted mean and standard deviation, which we report in Table \ref{tab:sample}, and summarize in Table \ref{tab:youth}. For further information about interesting objects and trends in these line measurements, see Section \ref{sec:discussion}.

\section{Results}
\label{sec:results}

\subsection{H$\alpha$ (6563\AA) activity indicator}
\begin{figure*}	
\centering
\includegraphics[width=0.8\textwidth]{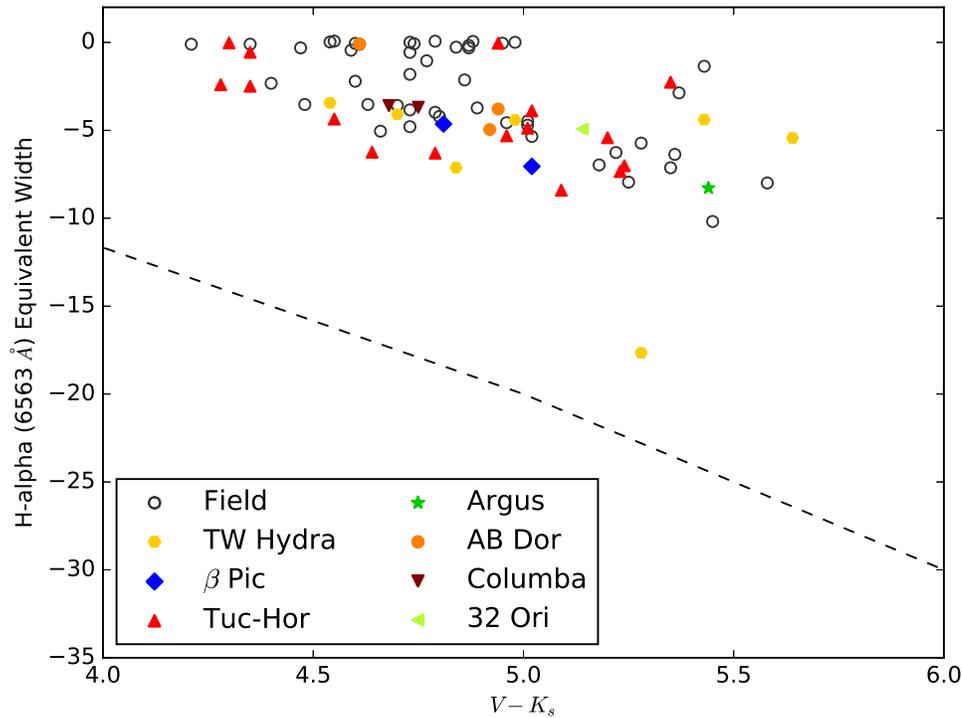}
\caption{H$\alpha$ equivalent widths versus $V-K$ color (as a proxy for spectral type), with veiling criterion (dashed line) from \citet{White2003}. All of these stars are post-T-Tauri stars, and the single star (represented by both measurements) that comes closest to the limit is TWA 3ABCD.}
\centering
\label{fig:Ha_VK}
\end{figure*}

H$\alpha$ emission, while related to youth, is not a reliable indicator of youth for M dwarfs as their activity can persist for billions of years. As shown by \citet{West2008}, M0-M2 stars with H$\alpha$ emission are typically younger than 1 Gyr; M3 stars are generally less than 2 Gyr old, suggesting most of our stars are younger than field age. Checking the T Tauri veiling limit in H$\alpha$ EW from \citet{White2003} suggests that none of the stars in this sample are potential T Tauri stars, which should be expected from the relative rarity of such objects within 100 parsecs, although RX~1924-3442 and TWA~3ABCD come close (Figure \ref{fig:Ha_VK}) with H$\alpha$ emission strengths greater than $-$10\AA, which are recorded in Table \ref{tab:youth}.

\subsection{Lithium (6708\AA) age indicator}
\begin{figure*}	
\centering
\includegraphics[width=0.8\textwidth]{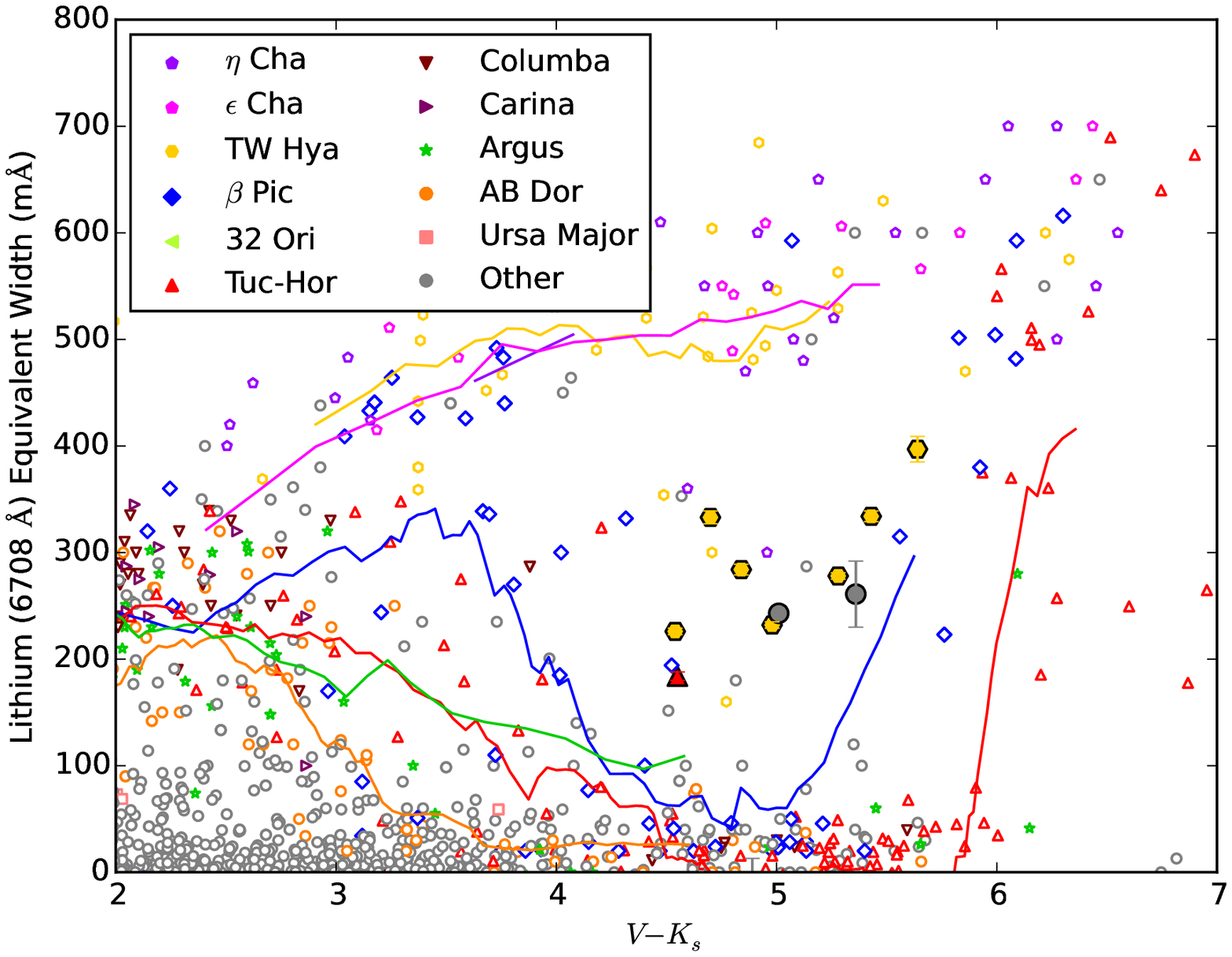}
\caption{Lithium equivalent width versus $V-K$. Ten stars in our sample with measurable lithium are shown with shapes outlined in black. The smaller shapes are stars from the Catalog of Suspected Nearby Young Stars \citep{Riedel2016a}, with 15-element moving averages plotted as rough trendlines for each group.}
\centering
\label{fig:lithium_VK}
\end{figure*}

Only ten objects had measurable lithium, with a typical measurement precision of 0.18\AA. This is expected, given that the selection consists of stars close to the convective limit where lithium is fused very quickly (see Figure \ref{fig:lithium_VK}, where they are plotted against lithium-bearing stars from the Catalog of Nearby Suspected Young Stars from \citealt{Riedel2016a}). With only three exceptions (a Tuc-Hor member and two stars that do not match any known groups), our ten lithium-detected stars are known or new members of the $\sim$12-Myr-old TW Hydra moving group, as outlined in Table \ref{tab:youth}. To define typical values for the groups, we computed a 15-element moving average and a 15-element moving standard deviation for each group, such that the value (and standard deviation) of each point on the curves are the average of the surrounding 15 points, and are given for the mean $V-K$ values of those 15 points. Interestingly, all of our detections lie below the curve for TW Hya members ($\sim$10 Myr) and above the curve for $\beta$ Pic ($\sim$25 Myr). Our targets have among the lowest lithium EWs measured for TW Hydra and the highest lithium EW for a Tuc-Hor member which suggests our unidentified young stars are under 25 Myr old. This is not entirely unexpected, given the apparent spread in lithium measurements shown in Figure \ref{fig:lithium_VK}.

\subsection{Potassium (7699\AA) gravity indicator}

\begin{figure*}	
\centering
\includegraphics[width=0.8\textwidth]{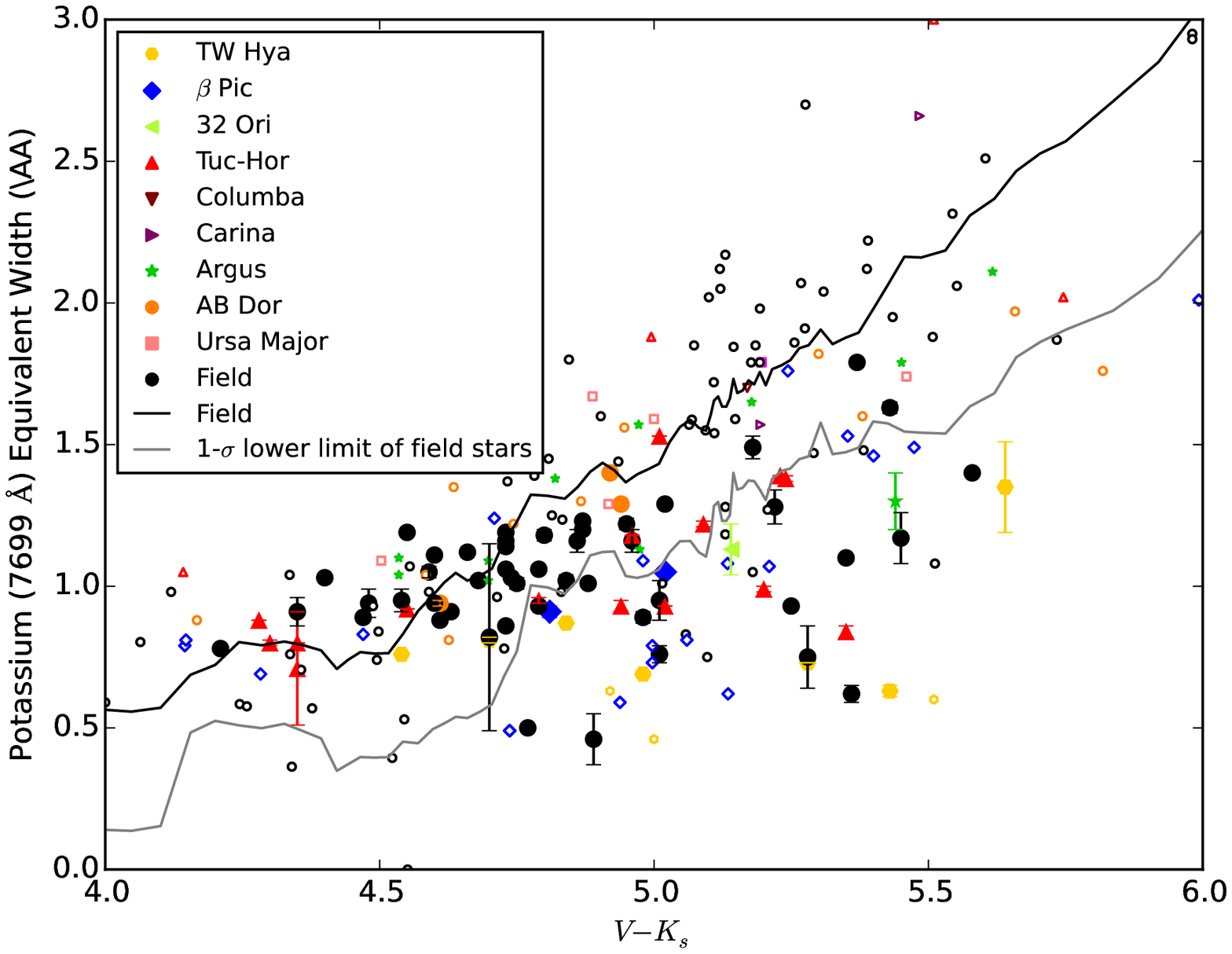}
\caption{Potassium equivalent width versus $V-K$. Stars appear according to Figure \ref{fig:lithium_VK}. The black curve is a 15-element windowed average fit to the field stars from the Catalog of Nearby Young Stars \citep{Riedel2016a} and the sample field stars in \citet{Riedel2014}. The main sequence locus has some width due to imprecise measurements in the CTIO 1.5m RCSpec data from \citet{Riedel2014}, and intrinsic scatter amongst group members. The gray line represents the 15-element windowed standard deviation of field stars and is roughly 0.5\AA. Stars that lie below the line (i.e. with weaker potassium absorption) are considered young.}
\centering
\label{fig:potassium}
\end{figure*}

The potassium 7699\AA~line\footnote{The other line in the doublet, potassium 7635\AA, is within the atmospheric A band and not suitable for measurement.} was measured for the entire sample, and results are shown in Figure \ref{fig:potassium}. In order to determine a standard of youth, the potassium EW measurements are compared to measurements of objects in the Catalog of Suspected Nearby Young Stars in \citet{Riedel2016a}, whose potassium measurements are largely from \citet{Riedel2014} and \citet{Shkolnik2009}, plus the additional field star measurements used in \citet{Riedel2014}. Both samples have relatively large uncertainties on the equivalent widths (0.2\AA~for \citealt{Riedel2014}, 0.24 or 0.16\AA~for \citealt{Shkolnik2009}, depending on the telescope) which, combined with intrinsic scatter, make the field star locus rather large (as defined by a 15-element moving average). Stars more than one standard deviation below the main sequence locus (as defined by a 15-element moving standard deviation) in Figure \ref{fig:potassium} are therefore treated as potentially low surface gravity objects and likely to be young, and this is used in the youth evaluation in Table \ref{tab:youth}. The lowest value, $-$0.07$\pm$0.01\AA, for SCR~2204-0711, is not shown in Figure \ref{fig:potassium} and suggests that the star is in fact a giant.

\subsection{Kinematic Results}

To supplement our radial velocities (Table \ref{tab:RVtable}, Section \ref{sec:RVs}), ICRS positions and proper motions for these objects were obtained from the Fourth USNO Compiled Astrographic Catalog (UCAC4, \citealt{Zacharias2013}) and PPMXL \citep{Roeser2010} catalogs. These are given in Table \ref{tab:supplementary}.

We use the LocAting Constituent mEmbers In Nearby Groups (LACEwING,  \citealt{Riedel2016a}) moving group identification code to evaluate membership probabilities in the NYMGs. LACEwING calculates up to four metrics of membership by comparing the proper motions, parallaxes, radial velocities, and space positions of targets (depending on what data is available) against predictions computed for a member of the group at that RA and DEC. These metrics are combined into a single goodness-of-fit value, and translated into a membership probability using the pre-calculated results of a simulation of 8 million stars.

Membership probabilities are calculated by taking the simulated stars and matching them to each of thirteen nearby young moving groups and three open clusters (Table \ref{tab:LACEwINGgroups}), and then computing a membership probability based on the result of combining four goodness-of-fit scores. Therefore, for a group X, there is a histogram of all the simulated stars with their goodness-of-fit scores when matched to group X. The histogram records the percentage of stars in each bin that were actually members of group X, effectively making the LACEwING percentages contamination probabilities. LACEwING does not force all probabilities to add up to 100\%, so there is still a chance that the stars will not match any known group, or add up to more than 100\% if the uncertainties are larger than the simulation expected.

\begin{deluxetable}{lrl}
\setlength{\tabcolsep}{0.02in}
\tablewidth{0pt}
\tabletypesize{\scriptsize}
\tablecaption{Moving Groups and Open Clusters Considered by the LACEwING Code\label{tab:LACEwINGgroups}}
\tablehead{
  \colhead{Name} &
  \colhead{Age}  &
  \colhead{Ref} \\
  \colhead{} &
  \colhead{(Myr)} &
  \colhead{} }
\startdata
$\epsilon$ Chameleontis & 5 & \citet{Murphy2013} \\
$\eta$ Chameleontis\tablenotemark{a} & 10 & \citet{Murphy2013} \\
TW Hydra & 10 & \citet{Weinberger2013} \\
32 Orionis & 20 & \citet{Bell2015} \\
$\beta$ Pictoris & 25 & \citet{Bell2015} \\
Octans & 40 & \citet{Murphy2015} \\
Tucana-Horologium & 45 & \citet{Bell2015} \\
Columba & 45 & \citet{Bell2015} \\
Carina & 45 & \citet{Bell2015} \\
Argus & 50 & \citet{Barrado-y-Navascues2004} \\
AB Doradus & 150 & \citet{Bell2015} \\
Carina-Near & 200 & \citet{Zuckerman2006} \\
Ursa Major & 400 & \citet{Jones2015} \\
Coma Berenices\tablenotemark{a} & 400 & \citet{Kraus2007} \\
$\chi^1$ Fornax & 525 & \citet{Pohnl2010} \\
Hyades\tablenotemark{a} & 800 & \citet{Brandt2015} \\
\enddata
\tablenotetext{a}{Open Cluster}
\tablecomments{Ages have been rounded to the nearest 5 Myr.}
\end{deluxetable}

As shown in \citet{Riedel2016a}, kinematic identification of young stars improves with more and higher-precision data. By including radial velocities, we significantly decrease the false positive rate. However, because LACEwING uses standard deviations in its four membership metrics, the lower quality of radial velocities (5 km s$^{-1}$) compared to the ones LACEwING is calibrated for (1 km s$^{-1}$) will result in better apparent matches and higher membership probabilities than would otherwise be expected.

LACEwING has two modes: one including a field star population 50 times the size of the NYMG population, implicitly assuming that the star under consideration might be a field interloper with coincidentally similar motions; and one for use if the star is already known to be young, where nearly all of the field star population is removed from consideration, leaving only a 1:1 contribution of field stars:NYMG members to represent the fact that \citet{Riedel2016a} found half of all young stars (there, defined as lithium-rich objects) were not members of any NYMG (see also Section \ref{sec:conclusions}), suggesting the presence of a young field. Stars with lithium detections or with potassium line strengths more than 1-$\sigma$ weaker than field stars were assumed to be young. All other stars were run through LACEwING's field star mode.

For 13 stars with parallaxes (mostly from \citealt{Riedel2014}), we can also compute full UVW space velocities and space positions following the matrix method of \citet{Johnson1987}. These are given in Table \ref{tab:UVWs}.

\begin{deluxetable*}{lrrrrrr}
\setlength{\tabcolsep}{0.02in}
\tablewidth{0pt}
\tabletypesize{\scriptsize}
\tablecaption{$UVWXYZ$ space velocities and positions\label{tab:UVWs}}
\tablehead{
  \colhead{Name}   &
  \colhead{$U$}    &
  \colhead{$V$}    &
  \colhead{$W$}    &
  \colhead{$X$}    &
  \colhead{$Y$}    &
  \colhead{$Z$}    \\
  \colhead{}       &
  \colhead{(km s$^{-1}$)}    &
  \colhead{(km s$^{-1}$)}    &
  \colhead{(km s$^{-1}$)}    &
  \colhead{(pc)}    &
  \colhead{(pc)}    &
  \colhead{(pc)}    }
\startdata
SCR~0017-6645   & -17.7$\pm$2.5 &  -8.8$\pm$3.0 &   2.9$\pm$4.3 &  15.6$\pm$1.1 & -19.8$\pm$1.4 & -30.1$\pm$2.1 \\
GJ~2006A        & -11.6$\pm$0.8 & -14.9$\pm$1.0 &  24.1$\pm$4.6 &   4.1$\pm$0.2 &  -1.1$\pm$0.1 & -32.1$\pm$1.8 \\
GJ~2006B        & -12.2$\pm$1.3 & -13.4$\pm$1.2 &  -8.0$\pm$6.1 &   4.1$\pm$0.2 &  -1.1$\pm$0.1 & -32.1$\pm$1.8 \\
HIP~3556        & -11.3$\pm$1.7 & -18.7$\pm$2.7 &   4.8$\pm$4.5 &   9.8$\pm$1.1 & -13.9$\pm$1.5 & -37.2$\pm$4.1 \\
BAR~161-12      & -13.0$\pm$2.0 &  -9.8$\pm$1.1 & -17.7$\pm$5.3 & -10.1$\pm$0.1 &   5.2$\pm$0.0 & -27.4$\pm$0.2 \\
SCR~1012-3124AB & -14.7$\pm$1.4 & -17.8$\pm$5.2 &  -7.8$\pm$2.3 &  -2.4$\pm$0.2 & -51.0$\pm$4.9 &  18.8$\pm$1.8 \\
SCR~1121-3845   & -13.9$\pm$1.0 & -16.6$\pm$2.0 &  -6.7$\pm$1.0 &  14.8$\pm$0.7 & -58.2$\pm$2.6 &  22.8$\pm$1.0 \\
TWA~5ABC        & -11.9$\pm$0.7 & -17.7$\pm$1.9 &  -5.9$\pm$1.0 &  11.6$\pm$0.4 & -43.8$\pm$1.5 &  21.6$\pm$0.8 \\
RX~1132-2651A   & -14.7$\pm$1.6 & -17.9$\pm$2.2 &  -8.1$\pm$1.9 &   8.2$\pm$0.4 & -38.7$\pm$1.9 &  25.5$\pm$1.2 \\
2MASS~1207-3247 & -18.4$\pm$1.1 &   2.1$\pm$2.4 & -17.5$\pm$1.5 &  17.9$\pm$0.5 & -43.6$\pm$1.1 &  26.3$\pm$0.7 \\
SCR~1425-4113AB &  -0.9$\pm$2.9 & -24.5$\pm$2.5 &  -4.6$\pm$1.4 &  49.8$\pm$3.2 & -39.9$\pm$2.6 &  21.1$\pm$1.4 \\
SCR~2010-2801AB &  -2.6$\pm$3.6 & -11.7$\pm$1.3 & -12.0$\pm$2.2 &  41.0$\pm$2.7 &  10.3$\pm$0.7 & -23.0$\pm$1.5 \\
L~755-19        & -28.9$\pm$4.4 & -20.7$\pm$2.9 &  -1.1$\pm$2.7 &  14.0$\pm$0.4 &   9.2$\pm$0.3 &  -8.5$\pm$0.3 \\
\enddata
\tablecomments{All values computed using right-handed coordinates, where U/X is toward the center of the Galaxy, V/Y in the direction of motion, and W/Z toward the north Galactic pole. Memberships for stars are given in Table \ref{tab:youth}}
\end{deluxetable*}

\section{Discussion}
\label{sec:discussion}

Our spectroscopic indicators are only sensitive to the youngest stars in this sample of hot M dwarfs. As shown in Figure \ref{fig:lithium_VK}, we should only expect to detect significant lithium in members of the youngest groups: $\epsilon$ Cha (5 Myr), the $\eta$ Cha open cluster (8 Myr), and TW Hya (10 Myr); our sample spans a range of temperatures where lithium is destroyed very quickly. None of our stars are young enough to meet the H$\alpha$ EW veiling criterion (Figure \ref{fig:Ha_VK}); all are post-T-Tauri stars. The relatively low precision of our potassium line measurements (Figure \ref{fig:potassium}) demonstrates that Tuc-Hor (45 Myr) is the oldest group for which significant numbers of stars fall below the field distribution. Any star that meets any of the spectroscopic criteria for youth is therefore definitely young, and most likely in Argus (50 Myr old) or the younger NYMGs (see Table \ref{tab:LACEwINGgroups}). Members of older NYMGs like AB Doradus (150 Myr old) should not meet our spectroscopic young criterion. These objects may still be identified by kinematics, though it is important to reiterate that kinematic techniques make no comment on the actual age of stars.

We take the most likely group reported by the LACEwING kinematic code as the correct membership, unless it is below 20\% probability, or the star is clearly younger than the most likely membership.

In total, we have identified 44 young systems (46 young stars) using the SALT spectroscopy, as shown in Table \ref{tab:youth}.
\begin{deluxetable*}{lllllrrrr}
\tabletypesize{\small}
\tablewidth{0pt}
\tablecolumns{9}
\tablecaption{Youth Properties\label{tab:youth}}
\tablehead{ 
\colhead{Name}   & 
\colhead{Youth}  &
\multicolumn{2}{c}{Literature} &
\multicolumn{2}{c}{LACEwING} &
\colhead{Kine. dist}  &
\colhead{Kine. RV}    &
\colhead{Measured RV} \\
\colhead{}            &
\colhead{Flags}       &
\colhead{Membership}  &
\colhead{Ref.}        &
\colhead{Membership}  &
\colhead{Prob.} &
\colhead{(pc)}        & 
\colhead{(km s$^{-1}$)}&
\colhead{(km s$^{-1}$)}
}
\startdata
SCR~0017-6645   & ~~K &$\beta$ Pic&  1 & Tuc-Hor   &  50 & 48.2$\pm$4.5 &  $+$7.4$\pm$3.5 &   $-$4.8$\pm$5.6 \\
GJ~2006A        & ~~K &$\beta$ Pic&  2 & $\beta$ Pic\tablenotemark{a} &   0 & 35.8$\pm$3.3 &  $+$8.3$\pm$1.7 &  $-$24.9$\pm$4.6 \\
GJ~2006B        & ~~K &$\beta$ Pic&  2 &$\beta$ Pic&  49 & 34.3$\pm$3.2 &  $+$8.3$\pm$1.6 &   $+$6.9$\pm$6.1 \\
HIP~3556        &     & Tuc-Hor   &  3 & Tuc-Hor   &  74 & 40.8$\pm$3.8 &  $+$5.7$\pm$3.4 &   $-$0.7$\pm$4.9 \\
SCR~0106-6346   &     &           &    & Tuc-Hor   &  25 & 39.2$\pm$4.9 &  $+$8.7$\pm$3.5 &  $+$13.3$\pm$5.2 \\
{[PS78]}~190    &     &           &    &           &     &              &                 &   $+$3.6$\pm$5.4 \\
BAR~161-12      & ~~K &$\beta$ Pic&  4 &           &     &              &                 &  $+$19.0$\pm$5.8 \\
GIC~138         &     &           &    &           &     &              &                 &  $-$26.4$\pm$7.2 \\
L~173-39        &     &           &    &           &     &              &                 &  $+$33.6$\pm$5.5 \\
SCR~0149-5411   &     &           &    & Tuc-Hor   &  22 & 37.1$\pm$4.9 &  $+$9.4$\pm$3.3 &   $+$2.1$\pm$4.1 \\
SCR~0152-5950   &     & Tuc-Hor   &  5 & Tuc-Hor   &  38 & 39.1$\pm$5.5 & $+$10.1$\pm$3.3 &  $+$14.0$\pm$5.3 \\
SCR~0212-5851   &     & Tuc-Hor   &  6 &           &     &              &                 &   $+$4.5$\pm$5.0 \\
SCR~0213-4654   &     &           &    &           &     &              &                 &   $+$9.4$\pm$6.0 \\
SCR~0215-0929   &     & Tuc-Hor   &  1 &           &     &              &                 &  $+$15.4$\pm$5.5 \\
SCR~0220-5823   &     & Tuc-Hor   &  6 & Tuc-Hor   &  27 & 43.4$\pm$6.8 & $+$11.3$\pm$3.3 &  $+$18.3$\pm$6.3 \\
SCR~0222-6022   & ~~K & Tuc-Hor   &  6 & Tuc-Hor\tablenotemark{a} &  42 & 30.7$\pm$4.9 & $+$11.5$\pm$3.3 &  $+$28.6$\pm$7.0 \\
2MASS~0236-5203 & L   & Tuc-Hor   &  7 & Tuc-Hor   &  95 & 41.8$\pm$6.3 & $+$11.7$\pm$3.1 &  $+$12.9$\pm$5.5 \\
LP~886-73       & ~~K &           &    &           &     &              &                 &  $-$16.5$\pm$6.4 \\
SCR~0248-3404   & ~~K &           &    & Tuc-Hor   &  87 & 46.2$\pm$5.4 & $+$11.0$\pm$2.7 &  $+$14.5$\pm$5.5 \\
SCR~0254-5746   &     &           &    &           &     &              &                 &   $+$0.1$\pm$5.0 \\
2MASS~0254-5108A&     & Tuc-Hor   &  7 & Tuc-Hor   &  37 & 44.0$\pm$6.9 & $+$12.6$\pm$3.1 &  $+$13.0$\pm$5.1 \\
SCR~0256-6343   &     & Tuc-Hor   &  8 &           &     &              &                 &  $-$10.9$\pm$6.0 \\
LP~831-35       &     &           &    & AB Dor    &  32 & 27.4$\pm$0.1 & $+$23.0$\pm$1.9 &  $+$25.5$\pm$6.1 \\
2MASS~0510-2340A&     & Columba   &  1 & Columba   &  27 & 50.2$\pm$4.8 & $+$23.4$\pm$1.9 &  $+$18.8$\pm$5.2 \\
2MASS~0510-2340B&     & Columba   &  1 & Columba   &  20 & 58.4$\pm$5.9 & $+$23.4$\pm$1.9 &  $+$15.6$\pm$7.5 \\
SCR~0522-0606   & ~~K &           &    & 32 Ori    &  59 & 87.6$\pm$6.5 & $+$21.1$\pm$0.4 &   $-$1.5$\pm$5.0 \\
SCR~0711-3510AB &     & AB Dor    &  1 &           &     &              &                 &   $+$4.1$\pm$5.0 \\
SCR~0844-0637   &     &           &    &           &     &              &                 &  $-$18.0$\pm$2.0 \\
LP~728-71       &     &           &    &           &     &              &                 &   $-$5.6$\pm$3.4 \\
SCR~1012-3124AB & LK  & TW Hya    &  2 & TW Hya    &  90 & 41.8$\pm$8.6 & $+$15.7$\pm$2.3 &  $+$14.6$\pm$5.5 \\
TWA~3ABCD       & LKh & TW Hya    &  9 & TW Hya    &  57 & 34.5$\pm$6.4 & $+$12.6$\pm$2.2 &  $+$14.5$\pm$2.5 \\
SCR~1121-3845   & L   & TW Hya    & 10 & TW Hya    & 100 & 56.0$\pm$10.2& $+$12.1$\pm$2.2 &   $+$9.5$\pm$2.2 \\
TWA~5ABC        & L   & TW Hya    & 11 & TW Hya    & 100 & 47.4$\pm$8.6 & $+$11.1$\pm$2.2 &  $+$10.2$\pm$2.2 \\
RX~1132-3019    & L   & TW Hya    & 12 & TW Hya    &  78 & 43.6$\pm$7.9 & $+$10.7$\pm$2.1 &  $+$15.8$\pm$4.7 \\
RX~1132-2651A   & LK  & TW Hya    & 13 & TW Hya    &  88 & 38.6$\pm$7.6 & $+$10.3$\pm$2.1 &   $+$7.8$\pm$2.5 \\
SIPS~1145-4055  &     &           &    &           &     &              &                 &  $+$14.2$\pm$6.2 \\
LP~851-410      & ~~K &           &    &           &     &              &                 &  $-$36.7$\pm$4.3 \\
SCR~1200-1731   & LK  &           &    &           &     &              &                 &  $+$20.1$\pm$2.4 \\
2MASS~1207-3247 & ~~K & TW Hya    & 14 &           &     &              &                 &  $-$16.4$\pm$2.9 \\
L~758-107       & ~~K &           &    &           &     &              &                 &  $-$12.0$\pm$4.9 \\
SCR~1230-3300   &     &           &    &           &     &              &                 &  $+$20.7$\pm$2.2 \\
SCR~1233-3641   &     &           &    &           &     &              &                 &   $+$6.7$\pm$4.4 \\
SCR~1237-4021   & LK  &           &    & TW Hya    &  88 & 61.4$\pm$10.6&  $+$7.6$\pm$2.3 &  $+$11.3$\pm$2.4 \\
SCR~1238-2703   &     & AB Dor    &  1 &           &     &              &                 &   $+$0.5$\pm$3.3 \\
SCR~1316-0858   &     &           &    &           &     &              &                 &   $+$6.1$\pm$4.9 \\
SCR~1321-1052   & ~~K &           &    &           &     &              &                 &  $+$26.5$\pm$5.6 \\
SCR~1421-0916   & ~~K &           &    &           &     &              &                 &   $-$4.3$\pm$4.3 \\
SCR~1421-0755   & ~~K &           &    &           &     &              &                 &   $-$5.2$\pm$7.2 \\
SCR~1425-4113AB & LK  & TW Hya    &  2 &           &     &              &                 &  $+$12.4$\pm$3.7 \\
SCR~1438-3941   &     &           &    &           &     &              &                 &  $+$36.6$\pm$5.9 \\
LP~914-6        &     &           &    &           &     &              &                 &  $+$17.5$\pm$4.1 \\
SCR~1521-2514   &     &           &    &           &     &              &                 &   $-$0.8$\pm$2.9 \\
SCR~1708-6936   &     & Tuc-Hor   &  1 &           &     &              &                 &  $+$11.3$\pm$3.4 \\
SCR~1816-6305   &     &           &    &           &     &              &                 &  $+$33.5$\pm$0.9 \\
SCR~1842-5554A  & ~~K &$\beta$ Pic&  1 &           &     &              &                 &   $+$9.8$\pm$4.9 \\
NLTT~47004AB    &     &           &    &           &     &              &                 &  $+$21.0$\pm$4.0 \\
SCR~1856-6922   &     &           &    &           &     &              &                 &  $+$21.1$\pm$4.0 \\
WT~625          &     &           &    &           &     &              &                 &  $-$15.6$\pm$4.3 \\
SCR~1922-6310   &     & Tuc-Hor   &  1 &           &     &              &                 &   $+$5.7$\pm$2.1 \\
RX~1924-3442    & ~~Kh&$\beta$ Pic&  1 &           &     &              &                 &   $-$7.7$\pm$5.7 \\
SCR~1926-5331   & ~~K &           &    & Tuc-Hor   &  42 & 57.2$\pm$7.5 &  $-$2.8$\pm$3.9 &   $-$7.6$\pm$6.0 \\
SCR~1938-2416   &     &           &    &           &     &              &                 &  $+$14.4$\pm$3.9 \\
SCR~1951-4025   &     &           &    &           &     &              &                 &  $+$20.6$\pm$5.9 \\
SCR~2004-6725A  &     &           &    &           &     &              &                 &  $+$10.4$\pm$3.5 \\
2MASS~2004-3356 & ~~K &$\beta$ Pic& 15 & Argus     &  44 & 30.6$\pm$0.7 & $-$18.4$\pm$1.8 &  $-$16.1$\pm$3.8 \\
SCR~2008-3519   & ~~K &           &    & Tuc-Hor   &  29 & 54.3$\pm$6.6 &  $-$9.8$\pm$4.0 &   $-$5.9$\pm$6.3 \\
SCR~2010-2801AB & ~~K &$\beta$ Pic&  1 &           &     &              &                 &   $+$1.0$\pm$4.2 \\
L~755-19        &     & Argus     &  2 &           &     &              &                 &  $-$31.2$\pm$5.8 \\
SCR~2107-7056   &     &           &    & Tuc-Hor   &  23 & 50.0$\pm$4.6 &  $+$4.6$\pm$3.8 &   $+$3.3$\pm$4.6 \\
SCR~2107-1304   &     &$\beta$ Pic&  1 &           &     &              &                 &   $-$3.7$\pm$4.8 \\
LEHPM~1-4147    &     &           &    &           &     &              &                 &  $+$12.4$\pm$4.0 \\
SCR~2204-0711   & ~~K &           &    &           &     &              &                 & \\
SCR~2237-2622   & ~~K & Argus     &  1 & Tuc-Hor   &  33 & 44.8$\pm$0.5 &  $-$8.2$\pm$3.5 &   $+$3.4$\pm$5.0 \\
SIPS~2258-1104  &     &           &    &           &     &              &                 &  $+$20.2$\pm$5.5 \\
LEHPM~1-5404    &     &           &    & AB Dor    &  32 & 21.5$\pm$0.2 & $+$15.0$\pm$1.9 &  $+$12.3$\pm$6.7 \\
SCR~2328-6802   &     & Tuc-Hor   &  1 & Tuc-Hor   &  32 & 49.6$\pm$2.5 &  $+$6.3$\pm$3.7 &   $+$3.5$\pm$5.1 \\
LTT~9582        &     & AB Dor    &  1 & AB Dor    &  31 & 23.3$\pm$0.1 & $+$12.1$\pm$1.9 &   $+$8.7$\pm$5.3 \\
G~275-71        & ~~K &           &    & Tuc-Hor\tablenotemark{a} &  48 & 29.7$\pm$0.9 & $-$5.1$\pm$3.1 &   $-$2.9$\pm$7.7 \\
LEHPM~1-6053    &     &           &    &           &     &              &                 &  $+$25.2$\pm$5.7 \\
\enddata

\tablecomments{Membership flags are based on spectroscopic analysis (see Section \ref{sec:results}) and are given as follows: (K) potassium EW below standard deviation of field value, (L) detectable lithium absorption, (h) H$\alpha$ emission stronger than $-$10\AA. Values are given in Table \ref{tab:sample}. Measured radial velocities are reprinted from Table \ref{tab:RVtable}. Membership references are to the first paper that identified the object as a member, and are 1.) \citet{Malo2013}, 2.) \citet{Riedel2014}, 3.) \citet{Zuckerman2001}, 4.) \citet{Shkolnik2012}, 5.) \citet{Kiss2011}, 6.) \citet{Rodriguez2013}, 7.) \citet{Torres2000}, 8.) \citet{Kraus2014a}, 9.) \citet{de-la-Reza1989} 10.) \citet{Sterzik1999}, 11.) \citet{Gregorio-Hetem1992}, 12.) \citet{Looper2010}, 13.) \citet{Webb1999}, 14.) \citet{Song2003}, 15.) \citet{Gagne2015}}
\tablenotetext{a}{Initial match was to a different moving group; see Section \ref{sec:discussion} for details.}
\end{deluxetable*}

\subsection{TW Hya members (10 Myr)}
\label{sec:twhya}
A number of known TW Hya systems are in our sample, and we reproduce membership for all but 2MASS~1207-3247 (TWA 23), which is also our only member without a lithium detection.

\begin{figure*}	
\centering
\includegraphics[width=2.3in]{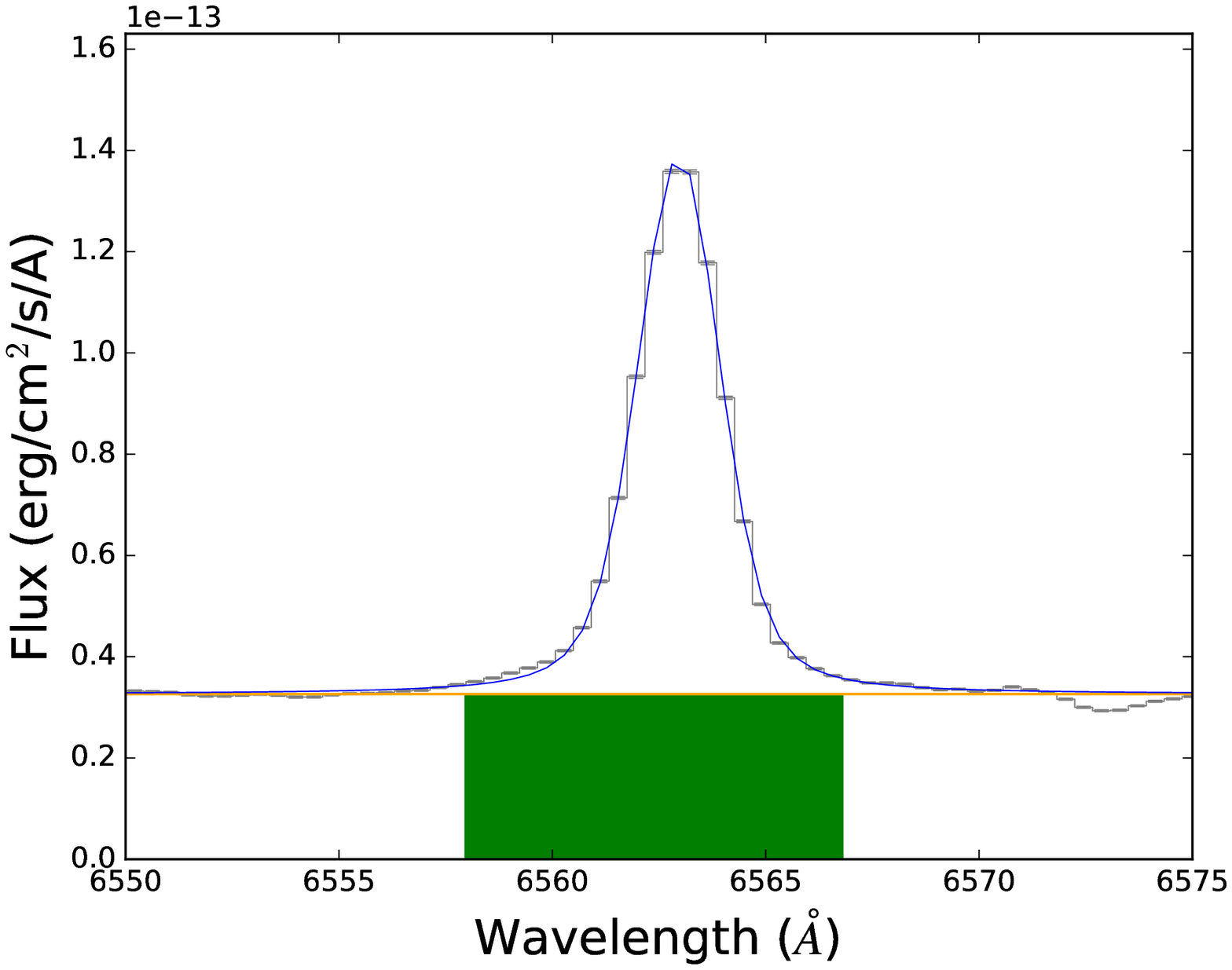}
\includegraphics[width=2.3in]{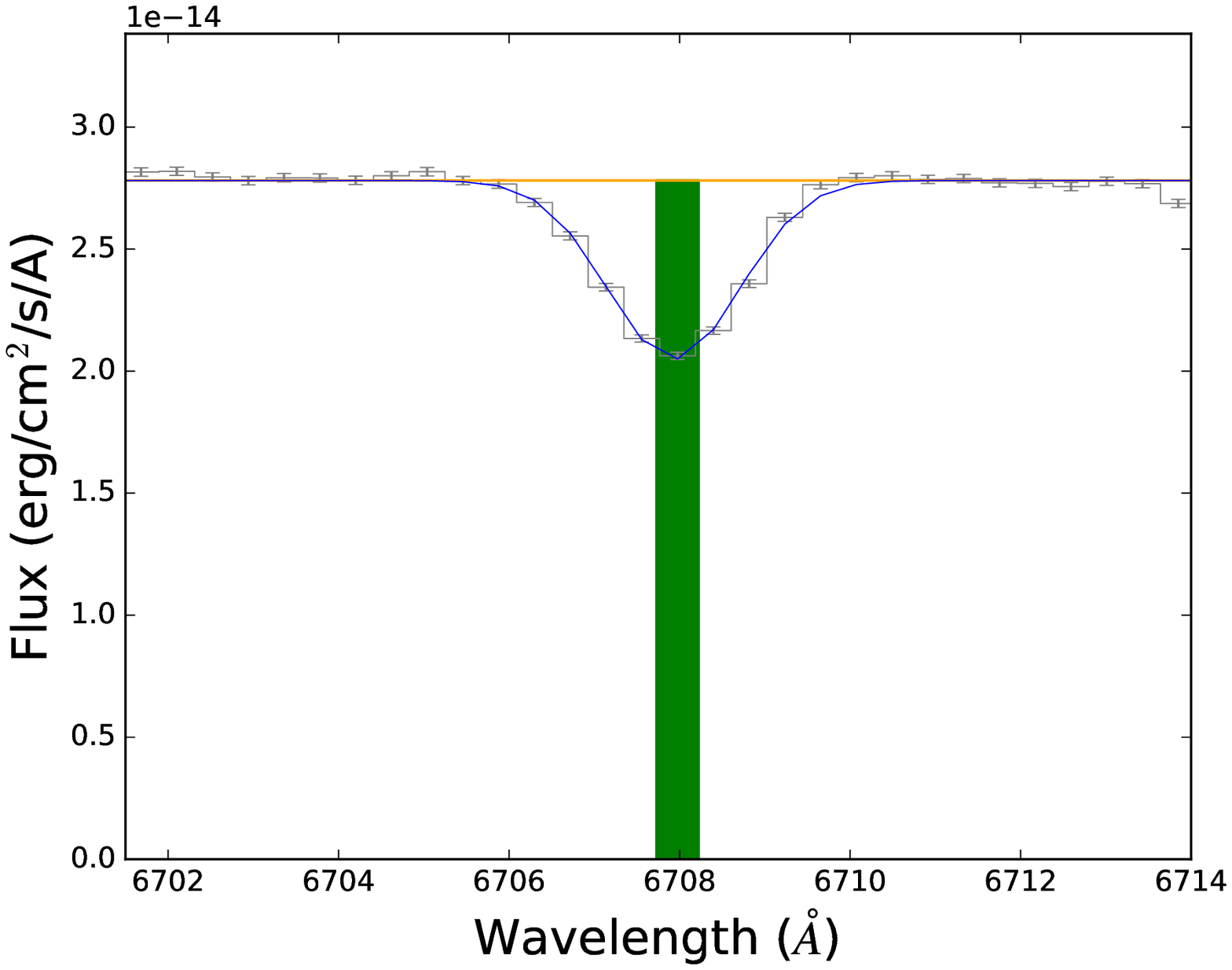}
\includegraphics[width=2.3in]{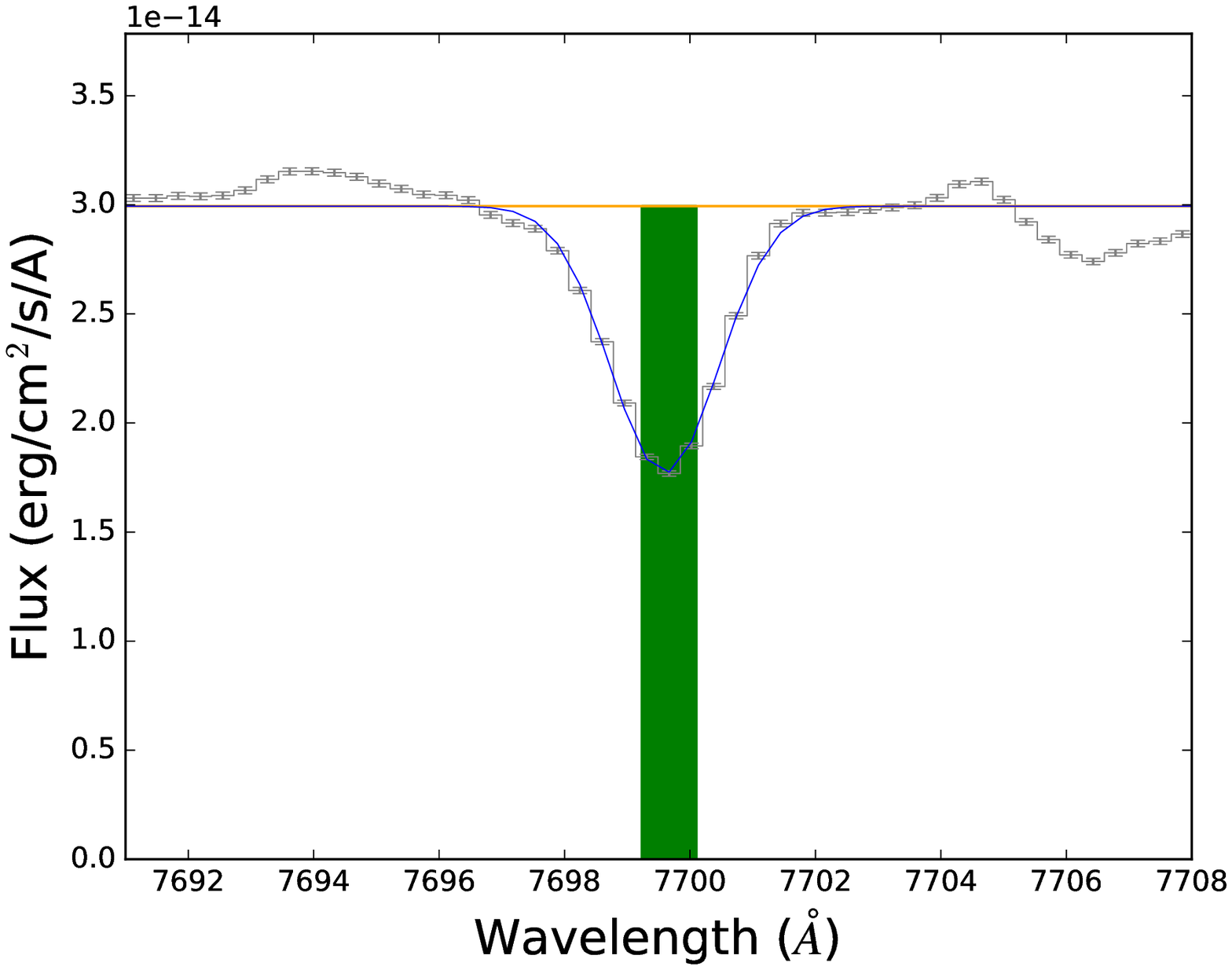}
\caption{The measured H$\alpha$ (6563 \AA) EW, Li (6708\AA) EW, and K {\sc i} (7699\AA) EW for SCR~1237-4021. The psuedocontinuum (yellow) and Voigt profile (blue) are fit to the observed spectrum (grey). The green rectangles approximate the EWs.}
\centering
\label{fig:SCR1237}
\end{figure*}

We have identified one new member of TW Hya. SCR~1237-4012 (Figure \ref{fig:SCR1237}) is confirmed by its lithium absorption, low surface gravity from potassium absorption, and kinematic match (88\%) to TW Hya.

SCR~1237-4012 is only 1957 arcseconds from TWA 11ABC, a triple system comprised of an A0 star, and two M2.5 companions. As an A0 star, TWA 11A is the most massive member of TW Hya. At the measured distance of TWA 11A (71.6$\pm$1.4 pc, consistent with the kinematic estimate of 61$\pm$10 pc), the projected separation of SCR~1237-4012 is only 0.68 parsecs.

Following the discussion of \citet{Mamajek2013} and \citet{Jiang2010}, we estimate the mass of TWA 11A at 2.3 solar masses\footnote{based on \url{http://www.pas.rochester.edu/~emamajek/spt/A0V.txt}, checked 20 Sep 2016} and TWA 11B and C at 0.4 solar masses each based on fits to 10 Myr \citet{Baraffe2010} isochrones. With a total system mass of 3 solar masses, the tidal radius should be 2 pc, and it is plausible that SCR~1237-4012 is an outer quadruple companion of the system. We note that given a population of 38 systems, within a 1$\sigma$ volume of 3000 cubic parsecs \citep{Riedel2016a}, the average separation between members of TW Hya should be on the order of 2.6 parsecs.

If we take the proper motion of TWA 11A (pmRA,pmDEC =  $-56.7\pm 0.3$, $-25.0\pm 0.2$ mas yr$^{-1}$ from \citealt{van-Leeuwen2007}) and SCR~1237-4012 (pmRA,pmDEC = $-63.7\pm 1.1$,$-29.1\pm 1.1$ mas yr$^{-1}$ from \citealt{Zacharias2013}), the proper motions differ by 8 mas yr$^{-1}$ (4$\sigma$), or 2.75 km s$^{-1}$ transverse velocity if they are at the same distance. There are two RVs for TWA 11: $7.1\pm 1.1$ km s$^{-1}$ from \citet{Gontcharov2006}, and $9.4\pm 2.3$ km s$^{-1}$ from \citet{Kharchenko2007}. If we compare the RVs to our RV for SCR~1237-4012, $11.3\pm 2.4$ km s$^{-1}$, the two systems' space velocities (if at the same distance) are separated by either 5.0 km s$^{-1}$ or 3.3 km s$^{-1}$. It is therefore somewhat less likely that the stars are in the same system, and more likely that they are either a chance alignment or that SCR~1237-4012 was ejected from the system within the past half million years.

A parallax measurement (and proper motion on a uniform system) of SCR~1237-4012 will do a great deal to determine if the 3D separation and space motion of SCR~1237-4012 is reasonable to make it a genuine companion to TWA 11A. 

\subsection{32 Ori member (20 Myr)}
The 32 Orionis moving group is relatively unstudied, and currently has only 17 known or suspected members. We have found another potential member, SCR~0522-0606, a low-surface-gravity M dwarf based on its potassium EW. The radial velocity is a 4$\sigma$ mismatch with LACEwING's expectations for a member of 32 Ori, but the quality of the proper motion agreement and resulting estimated spatial position still gives the star a 59\% probability of membership.

With its M2.5Ve spectral type, only three known members of the group have lower masses than SCR~0522-0606: two M3V stars, 2MASS J05253253+0625336 and 2MASS J05194398+0535021 \citep{Bell2015}, and one L1 low gravity brown dwarf \citep{Burgasser2016}. This is thus a useful benchmark object for the lower-mass end of the system.

\begin{figure}
\centering
\includegraphics[width=0.5\textwidth]{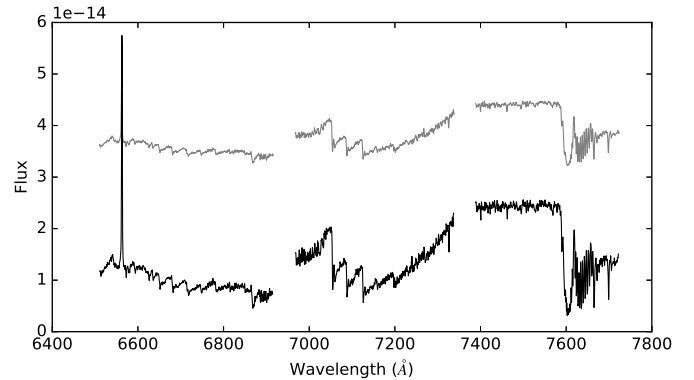}
\caption{Spectra of GJ 2006A (top, gray) and GJ 2006B (bottom, black).}
\centering
\label{fig:gj2006ab}
\end{figure}

\subsection{$\beta$ Pic member (25 Myr)}
The GJ~2006 system is a common proper motion visual binary that \citet{Riedel2014} identified as $\beta$ Pic members. For unknown reasons, our radial velocity for the A component is discrepant with both the B component and membership in $\beta$ Pic, regardless of the spectral region fit or combination of comparison stars. It is not clear why this velocity is different, as both components appear to be M dwarfs with H$\alpha$ in emission represented here by high-SNR spectra (Figure \ref{fig:gj2006ab}); further study is needed. For our purposes, we presume this velocity is in error, and identify both objects as $\beta$ Pic members in Table \ref{tab:youth}.

\subsection{Tuc-Hor members (45 Myr)}
2MASS~0236-5203 was one of the first identified members of Tuc-Hor \citep{Zuckerman2001}, and we reproduce that membership here. That it has a measurable lithium absorption feature is surprising given the age of Tuc-Hor, although not without precendent, as can be seen in Figure \ref{fig:lithium_VK}, and by existing lithium measurements of 2MASS~0236-5203 from \citet{Torres2006} and \citet{Mentuch2008}. It has a 23\% probability of membership in the younger $\beta$ Pic moving group, but we see no concrete reason to prefer that over Tuc-Hor membership.

SCR~0222-6022 has been considered as a member of Tuc-Hor before, by \citet{Rodriguez2013}, \citet{Kraus2014a}, and \citet{Malo2014a}. The first two papers rejected that membership. With our new radial velocity, we find a 93\% chance of membership in AB Dor, and only 42\% in Tuc-Hor. However, SCR~0222-6022 has low surface gravity according to the potassium feature, which indicates it is more likely a younger star and a member of the younger Tuc-Hor NYMG.

In addition to recovering seven known members, we identify nine new members of Tuc-Hor:

SCR~0017-6645 (50\%) was previously identified as a member of $\beta$ Pic by \citet{Malo2013} and \citet{Riedel2014}. We find only a 19\% probability that it is a member of $\beta$ Pic using our new radial velocity ($-$4.8$\pm$5.6 km s$^{-1}$). If we take a weighted mean of our radial velocity with the radial velocity from \citet{Malo2013} (11.4$\pm$0.8 km s$^{-1}$), the star is still more likely to be a member of Tuc-Hor. Given that this was one of the stars we removed from our fitting process in Section \ref{sec:RVs}, it may be that our radial velocity is in error. However, if we use only Malo's radial velocity, the star has only a marginally higher probability (73\%) of membership in $\beta$ Pic than Tuc-Hor (68\%).

Similarly, G~275-71 matches best to AB Dor (73\%), but given its low surface gravity, it is more likely to be a member of Tuc-Hor (48\%).

SCR~2237-2622 was previously identified by \citet{Malo2013} as a member of the Argus moving group. LACEwING results indicate a probability of 33\% in Tuc-Hor, 32\% in AB Dor, and no probability of membership in Argus.

SCR~0106-6346, SCR~0149-5411, SCR~0248-3404, SCR~1926-5331, SCR~2008-3519, and SCR~2107-7056 have never previously been identified as young stars. We note that SCR~0248-3404 was rejected as a member of Tuc-Hor by \citet{Kraus2014a}, and identified as an ambiguous nonmember by \citealt{Malo2013}; their published radial velocities $+$23.3$\pm$0.5 and $+$14.6$\pm$0.3 km s$^{-1}$ (respectively) differ by nearly 11$\sigma$. \citet{Malo2014a} found the system to be a single-lined spectroscopic binary (SB1). Our result ($+$14.5$\pm$5.5 km s$^{-1}$) is closer to that of \citet{Malo2013}.

\subsection{Columba members (45 Myr)}

We find one member of the Columba moving group: 2MASS~0510-2340AB. It was identified as such by \citet{Malo2013} and we reproduce that membership for both components. Despite being the fainter star of the pair, the B component has an earlier spectral type both here and in \citet{Malo2013}. This would suggest that they are actually two separate star systems, but their proper motions, kinematic distances, and radial velocities are identical at the 1-$\sigma$ level, and they are separated by only 27 arcsec. More study of this system is needed.

\subsection{Argus members (50 Myr)}
2MASS~2004-3356 was identified as an X-ray active (and therefore probably young) star by \citet{Riaz2006} and \citet{Haakonsen2009}, and as a member of $\beta$ Pic by \citet{Gagne2015}. Our radial velocity ($-$16 km s$^{-1}$) is inconsistent with membership in $\beta$ Pic, where $-$6 km s$^{-1}$ would be expected, and it has only a 14\% probability of membership in $\beta$ Pic according to LACEwING. With its low surface gravity and 44\% probability of membership in Argus, it is the only new member of Argus we have identified in this survey.

\subsection{AB Dor members (150 Myr)}
Two of the members of AB Dor identified here are new. LP 831-35 and LEHPM~1-5404 are not known as low gravity objects, which is consistent with their age.

\subsection{Young Nonmembers}
Perhaps most surprisingly, 14 of our stars are young -- under 50 Myr old as inferred from their potassium measurements -- but are not members of any NYMG according to their kinematics.

\begin{figure*}
\centering
\includegraphics[width=2.3in]{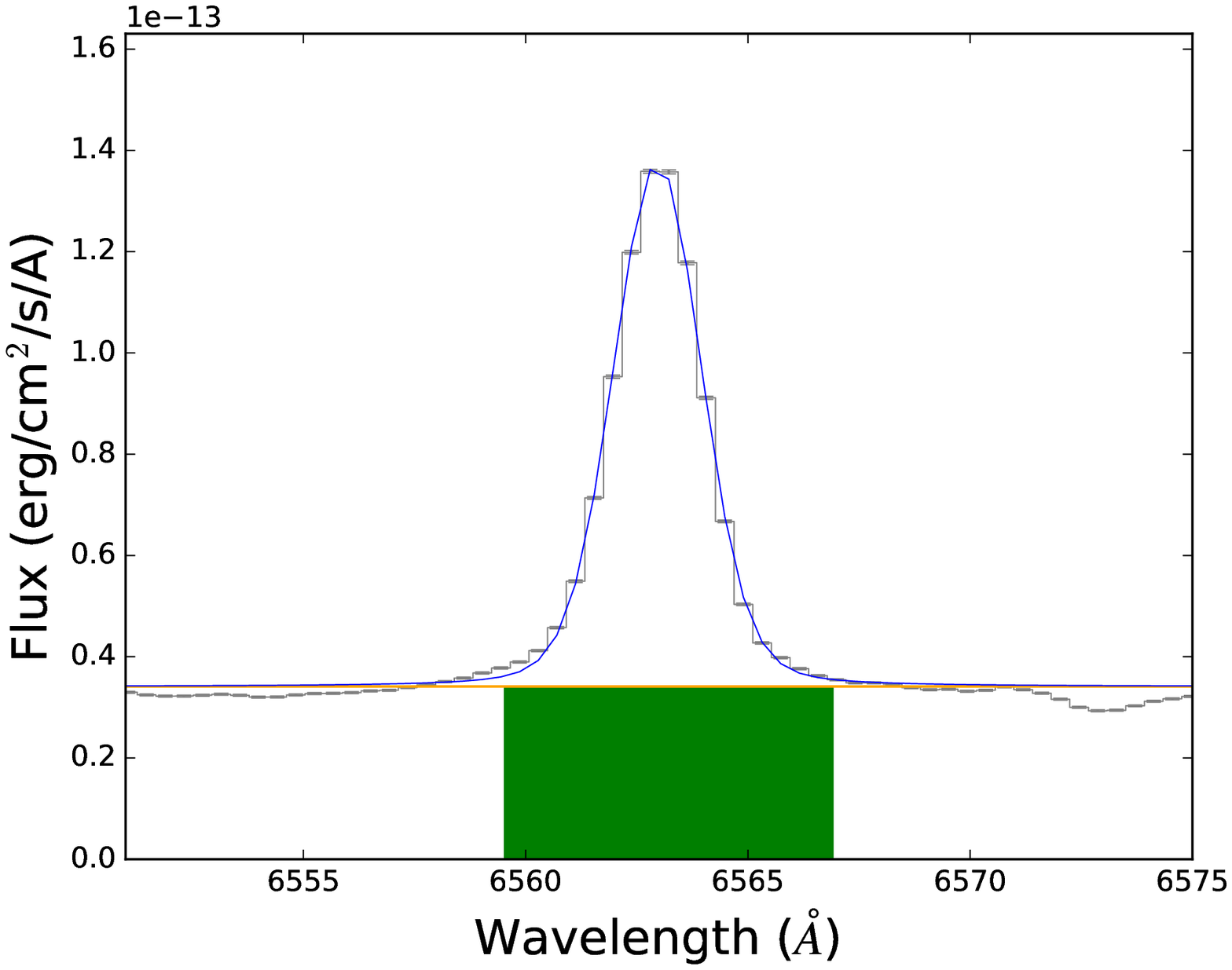}
\includegraphics[width=2.3in]{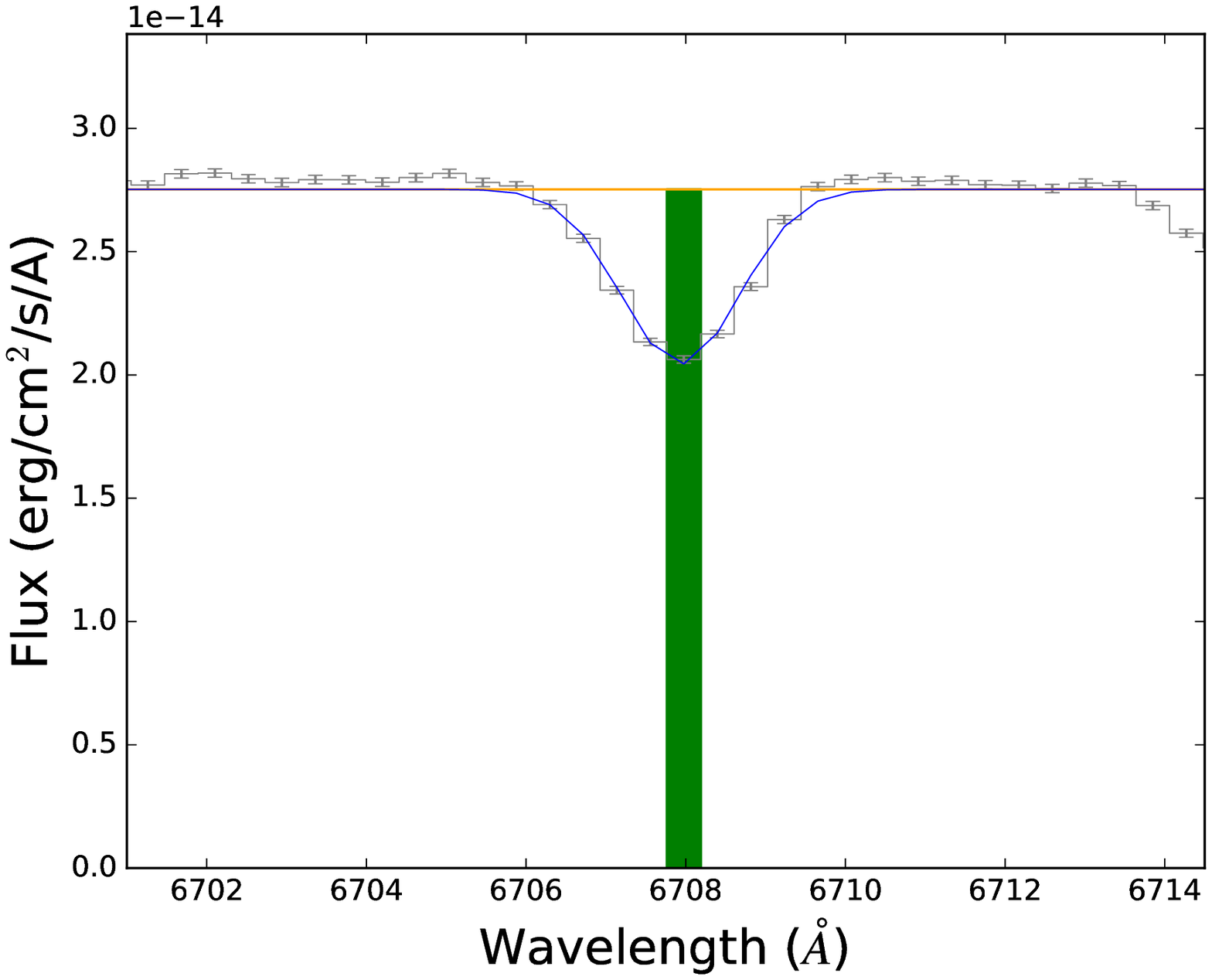}
\includegraphics[width=2.3in]{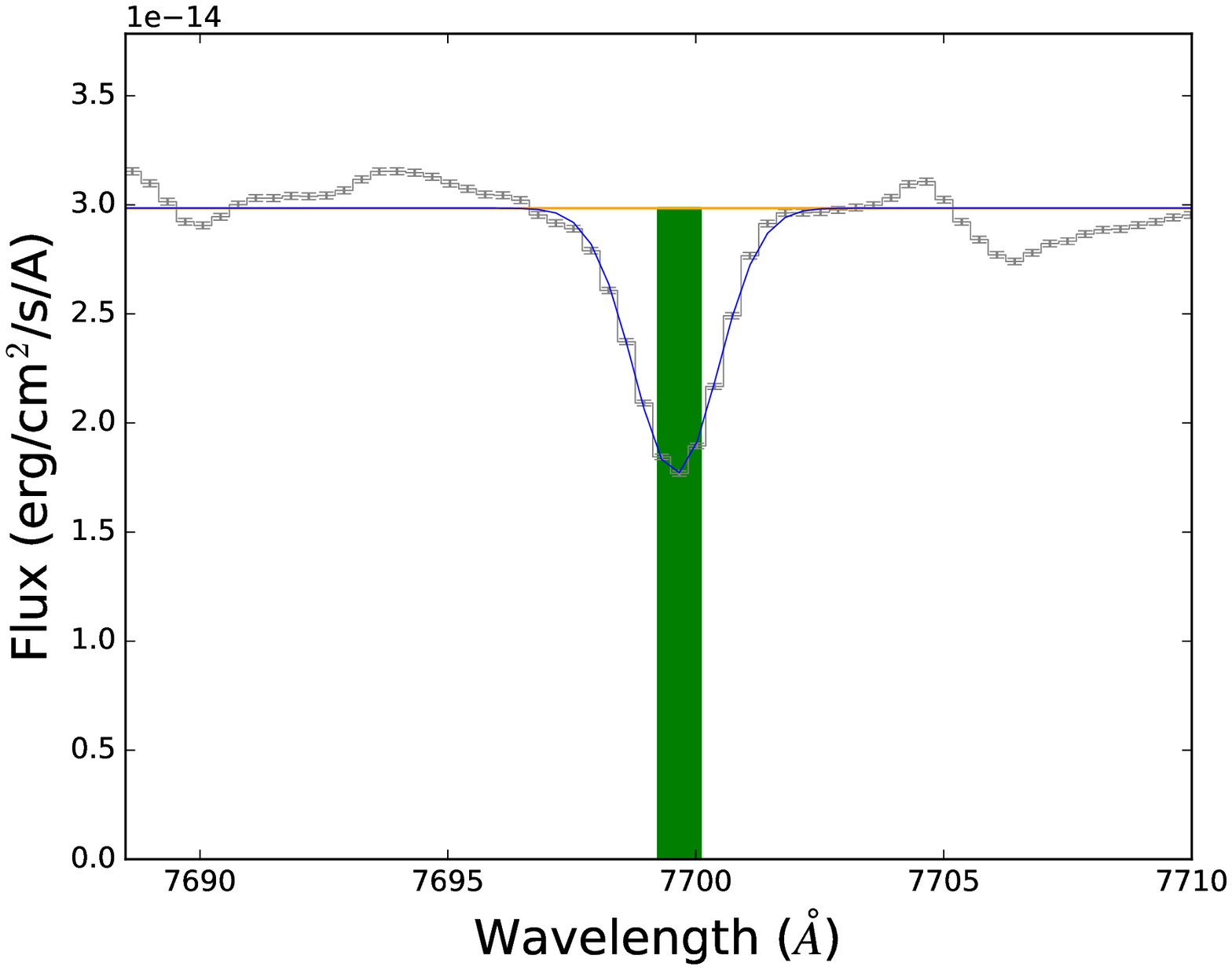}
\caption{The measured H$\alpha$ (6563 \AA) EW (top), Li (6708\AA) EW, and K {\sc i} (7699\AA) EW for SCR~1200-1731. The psuedocontinuum (yellow) and Voigt profile (blue) are fit to the observed spectrum (grey). The green rectangles approximate the EWs.}
\centering
\label{fig:SCR1200}
\end{figure*}

Two of these stars have lithium detections and must therefore be very young; most likely younger than Tuc-Hor (45 Myr): SCR~1425-4113AB and SCR~1200-1731. SCR~1425-4113AB was identified by \citet{Riedel2014} as a young star with potential kinematic matches to $\beta$ Pic and TW Hya. Based on its overluminosity on a color-magnitude diagram that placed it brighter than the TW Hya isochrone, it was determined more likely to be a TW Hya member despite its distance from the rest of the TW Hya members. \citet{Malo2014a} disputed this, placing the system in $\beta$ Pic. With our new radial velocity, we find no probability of membership in either group, although the star remains under 25 Myr old according to its spectra. SCR~1200-1731, despite a spatial location and lithium measurement (Figure \ref{fig:SCR1200}) that would likely make it a TW Hya member, has a radial velocity ($+$20.1 km s$^{-1}$) that disqualifies it entirely from membership in that group.

As mentioned in Section \ref{sec:twhya}, we did not reproduce the membership of 2MASS~1207-3247 in TW Hya. Our radial velocity for 2MASS~1207-3247 ($-$16.4$\pm$2.9) is very discrepant with the previously reported radial velocity $+$8.5$\pm$1.2 \citep{Malo2013}. If we combine that measurement with ours into a weighted mean ($+$4.8$\pm$8.8), we find a 100\% probability of membership in TW Hya. If we remove that radial velocity, we find that the proper motions on their own are also discrepant with membership in TW Hya. It is worth noting that this was one of the stars we removed from our RV comparisons in \ref{sec:RVs}.

BAR~161-12, SCR~1842-5554A, RX~1924-3442, and SCR~2010-2801AB have previously been identified as $\beta$ Pic members, but are not members based on our new radial velocities. All of them (except BAR~161-12, which matches nothing) still match $\beta$ Pic, though with probabilities too low to be meaningful.

L~755-19 was identified by \citet{Riedel2014} as a potential member of Argus, but with RV has a membership probability of only 19\%. 

LP~886-73, LP~851-410, L~758-107, SCR~1321-1052, SCR~1421-0755, and SCR~1421-0916 have never been suspected of membership in any moving group before, and their only claim to youth is the fact that they have low surface gravity according to our potassium measurements. While these identifications might possibly be spurious, we believe that we have demonstrated this search methodology's ability to identify and recover genuinely young stars, and that these stars are therefore also young field stars.

\subsection{Giants}
\label{sec:giants}
The negative potassium EW (Table \ref{tab:sample}) measured for SCR~2204-0711 (as well as our H$\alpha$ EW, and large and uncertain radial velocity in Table \ref{tab:RVtable}) is actually the result of the measurement code seeking out a completely different feature. This object appears to have NO measurable potassium absorption, and is likely to be a giant or supergiant, not a young star. This object has a spectral type of K9 but the colors of an M1-M6 dwarf.

\section{Conclusions}
\label{sec:conclusions}

We have obtained low-resolution red optical spectroscopy of 79 potentially nearby M dwarfs in 77 star systems. With that spectroscopy, we measured an age indicator, Li 6708\AA; a gravity indicator, K {\sc I} 7699\AA; an activity indicator, H$\alpha$ 6563\AA; and 5 km s$^{-1}$ precision radial velocities for every star in our sample. 

Using that information, we have identified 44 young star systems: seven members of TW Hydra, one member of $\beta$~Pic, one member of 32 Ori, 16 members of Tuc-Hor, one member of Columba, one member of Argus, three members of AB~Dor, 14 young systems that we cannot place among the nearby young moving groups; by adding radial velocities and spectroscopic confirmation, we reinforce the strength of the membership identifications. Of the young systems, 12 are new moving group members, including one new member of TW~Hya, one new member of 32 Ori, nine new members of Tuc-Hor, one new member of Argus, and two new members of AB~Dor. We have also discovered one giant. This study proves that the selection cuts imposed by the TINYMO survey are extremely effective at identifying young stars, considering the ratio of field stars to young stars is 25:1 \citep{Riedel2016a} in the Solar Neighborhood.

We find that the new TW Hya member, SCR~1237-4012, is potentially within the tidal radius of the triple-star system TWA 11, and may be a fourth member or recently ejected object. We have also identified nine new members of Tuc-Hor, making the numerically largest known moving group (209 systems, \citealt{Riedel2016a}; now 218) even larger. And we have added the third-lowest-mass member of the poorly explored 32 Ori moving group.

We also find further confirmation that a large number of young systems are not identifiably members of any known NYMGs -- 32\% (30\% if we allow for 2MASS~1207-3247 as an actual member of TW Hya) of our young star systems are not members of any NYMG. This has been noticed before. \citet{Riedel2016a} used a catalog constructed from NYMG membership papers, selected a sample of spectroscopically young (defined as having detectable lithium) and high confidence ``bona-fide'' members, and found that more than half of the systems were {\em not} members of any known NYMGs when analyzed uniformly with any publicly available moving group identification code. Our sample here is unlike the samples provided by most moving group identification papers. We did not start by kinematically identifying stars in a proper motion sample and then following up only likely candidates with spectroscopy; rather, our sample is biased only to bright low-proper-motion objects. Given that these stars genuinely appear to be young, further work is necessary to determine the true nature of these stars.

\section{Acknowledgements}
\acknowledgements

All of the observations reported in this paper were obtained with the Southern African Large Telescope (SALT), which is a partnership between the South African Astronomical Observatory and 11 international partners, under program codes 2014-1-AMNH-002 and 2014-2-AMNH-002. The generosity of the late Paul Newman and the Newman Foundation has made AMNH's participation in SALT possible. PyRAF is a product of the Space Telescope Science Institute, which is operated by AURA for NASA. ARR would like to thank Noel Richardson for help with flux calibration and RV calibration of the data, Jamie McDonald for editorial assistance, and Michael Shara for help with the data acquisition.

\software{
        astropy \citep{Astropy2013}, 
        LACEwING \citep{Riedel2016a},
        MATCHSTAR \citep{Riedel2014},
        PHEW \citep{PHEW},
        PyRAF, 
        PySpecKit \citep{Ginsburg2011}
}

\facility{SALT (RSS)}

\bibliographystyle{aasjournal} \bibliography{riedel_a}

\begin{deluxetable*}{lcrrrrrr}
\tabletypesize{\scriptsize}
\tablewidth{0pt}
\tablecolumns{8}
\tablecaption{SALT Equivalent Widths\label{tab:sample}}

\tablehead{ 
\colhead{Designated} & 
\colhead{Obs} &  
\multicolumn{2}{c}{H-$\alpha$ EW (\AA)} &
\multicolumn{2}{c}{Li EW (m\AA)} &
\multicolumn{2}{c}{K {\sc i} EW (\AA)} \\
\colhead{Name} &
\colhead{Date} &
\colhead{Measured} &
\colhead{Weighted} &
\colhead{Measured} &
\colhead{Weighted} &
\colhead{Measured} &
\colhead{Weighted} 
}
\startdata
SCR~0017-6645   & 2013-08-21 & -6.30 & -6.30$\pm$0.01 & \ldots & \ldots     & 0.96 & 0.95$\pm$0.01 \\
SCR~0017-6645   & 2013-08-21 & -6.30 &                & \ldots &            & 0.94 & \\
GJ~2006A        & 2013-08-18 & -4.63 & -4.64$\pm$0.01 & \ldots & \ldots     & 0.93 & 0.91$\pm$0.02 \\
GJ~2006A        & 2013-08-18 & -4.65 &                & \ldots &            & 0.88 & \\
GJ~2006B        & 2013-08-22 & -7.05 & -7.05$\pm$0.01 & \ldots & \ldots     & 1.06 & 1.05$\pm$0.01 \\
GJ~2006B        & 2013-08-22 & -7.06 &                & \ldots &            & 1.05 & \\
HIP~3556        & 2013-08-22 & -0.57 & -0.56$\pm$0.01 & \ldots & \ldots     & 0.80 & 0.80$\pm$0.01 \\
HIP~3556        & 2013-08-22 & -0.54 &                & \ldots &            & 0.79 & \\
SCR~0106-6346   & 2013-08-18 & -4.90 & -4.88$\pm$0.02 & \ldots & \ldots     & 1.53 & 1.53$\pm$0.01 \\
SCR~0106-6346   & 2013-08-18 & -4.86 &                & \ldots &            & 1.53 & \\
{[PS78]}~190    & 2013-08-22 & -0.57 & -0.57$\pm$0.01 & \ldots & \ldots     & 1.18 & 1.19$\pm$0.01 \\
{[PS78]}~190    & 2013-08-22 & -0.56 &                & \ldots &            & 1.20 & \\
BAR~161-12      & 2013-08-21 & -7.56 & -7.40$\pm$0.29 & \ldots & \ldots     & 1.10 & 1.10$\pm$0.01 \\
BAR~161-12      & 2013-08-21 & -6.89 &                & \ldots &            & 1.10 & \\
GIC~138         & 2014-01-17 & +0.06 & +0.08$\pm$0.01 & \ldots & \ldots     & 1.20 & 1.17$\pm$0.01 \\
GIC~138         & 2014-01-17 & +0.08 &                & \ldots &            & 1.17 & \\
L~173-39        & 2013-08-22 & -2.33 & -2.33$\pm$0.01 & \ldots & \ldots     & 1.03 & 1.03$\pm$0.01 \\
L~173-39        & 2013-08-22 & -2.32 &                & \ldots &            & 1.02 & \\
SCR~0149-5411   & 2014-01-18 & -0.03 & -0.03$\pm$0.01 & \ldots & \ldots     & 0.80 & 0.81$\pm$0.01 \\
SCR~0149-5411   & 2014-01-18 & -0.03 &                & \ldots &            & 0.81 & \\
SCR~0152-5950   & 2013-08-22 & -2.62 & -2.49$\pm$0.13 & \ldots & \ldots     & 0.51 & 0.71$\pm$0.20 \\
SCR~0152-5950   & 2013-08-22 & -2.36 &                & \ldots &            & 0.91 & \\
SCR~0212-5851   & 2013-08-20 & -3.60 & -3.54$\pm$0.06 & \ldots & \ldots     & 0.88 & 0.92$\pm$0.05 \\
SCR~0212-5851   & 2013-08-21 & -3.47 &                & \ldots &            & 0.98 & \\
SCR~0213-4654   & 2013-12-01 & -7.23 & -6.86$\pm$0.29 & \ldots & \ldots     & 1.45 & 1.49$\pm$0.04 \\
SCR~0213-4654   & 2013-12-01 & -6.64 &                & \ldots &            & 1.53 & \\
SCR~0215-0929   & 2013-08-23 & -5.06 & -5.05$\pm$0.01 & \ldots & \ldots     & 1.11 & 1.12$\pm$0.01 \\
SCR~0215-0929   & 2013-08-23 & -5.04 &                & \ldots &            & 1.13 & \\
SCR~0220-5823   & 2013-08-22 & -8.40 & -8.41$\pm$0.01 & \ldots & \ldots     & 1.23 & 1.22$\pm$0.01 \\
SCR~0220-5823   & 2013-08-22 & -8.41 &                & \ldots &            & 1.21 & \\
SCR~0222-6022   & 2013-08-23 & -7.33 & -7.34$\pm$0.01 & \ldots & \ldots     & 1.39 & 1.39$\pm$0.01 \\
SCR~0222-6022   & 2013-08-23 & -7.35 &                & \ldots &            & 1.39 & \\
2MASS~0236-5203 & 2013-08-22 & -4.36 & -4.36$\pm$0.01 & 180    & 184$\pm$4  & 0.92 & 0.92$\pm$0.01 \\
2MASS~0236-5203 & 2013-08-22 & -4.36 &                & 188    &            & 0.91 & \\
LP~886-73       & 2014-02-02 & -7.97 & -8.00$\pm$0.02 & \ldots & \ldots     & 1.40 & 1.40$\pm$0.01 \\
LP~886-73       & 2014-02-02 & -8.02 &                & \ldots &            & 1.41 & \\
SCR~0248-3404   & 2013-08-21 & -7.36 & -6.93$\pm$0.37 & \ldots & \ldots     & 1.38 & 1.38$\pm$0.01 \\
SCR~0248-3404   & 2013-08-21 & -6.61 &                & \ldots &            & 1.39 & \\
SCR~0254-5746   & 2013-08-21 & +0.04 & +0.02$\pm$0.02 & \ldots & \ldots     & 0.92 & 0.96$\pm$0.04 \\
SCR~0254-5746   & 2013-08-21 & +0.01 &                & \ldots &            & 0.99 & \\
2MASS~0254-5108A& 2013-08-22 & -2.47 & -2.40$\pm$0.06 & \ldots & \ldots     & 0.88 & 0.88$\pm$0.01 \\
2MASS~0254-5108A& 2013-08-23 & -2.34 &                & \ldots &            & 0.89 & \\
SCR~0256-6343   & 2013-08-21 & -6.29 & -6.26$\pm$0.03 & \ldots & \ldots     & 1.22 & 1.28$\pm$0.06 \\
SCR~0256-6343   & 2013-08-21 & -6.24 &                & \ldots &            & 1.33 & \\
LP~831-35       & 2013-08-23 & -4.88 & -4.95$\pm$0.07 & \ldots & \ldots     & 1.40 & 1.40$\pm$0.01 \\
LP~831-35       & 2013-08-23 & -5.02 &                & \ldots &            & 1.40 & \\
2MASS~0510-2340A& 2013-08-23 & -3.59 & -3.59$\pm$0.01 & \ldots & \ldots     & 1.02 & 1.02$\pm$0.01 \\
2MASS~0510-2340A& 2013-08-23 & -3.59 &                & \ldots &            & 1.03 & \\
2MASS~0510-2340B& 2014-02-27 & -3.54 &                & \ldots &            & 1.01 & \\                
2MASS~0510-2340B& 2013-10-23 & -3.84 & -3.73$\pm$0.10 & \ldots & \ldots     & 1.05 & 1.01$\pm$0.03 \\
2MASS~0510-2340B& 2013-10-23 & -3.76 &                & \ldots &            & 0.98 & \\
2MASS~0510-2340B& 2014-02-27 & -3.67 &                & \ldots &            & 1.01 & \\
SCR~0522-0606   & 2013-12-02 & -4.06 & -5.32$\pm$0.97 & \ldots & \ldots     & 1.05 & 1.15$\pm$0.09 \\
SCR~0522-0606   & 2013-12-02 & -6.06 &                & \ldots &            & 1.23 & \\
SCR~0711-3510AB & 2014-04-11 & -2.11 & -2.13$\pm$0.02 & \ldots & \ldots     & 1.20 & 1.16$\pm$0.04 \\
SCR~0711-3510AB & 2014-04-11 & -2.16 &                & \ldots &            & 1.13 & \\
SCR~0844-0637   & 2014-02-18 & -0.32 & -0.31$\pm$0.01 & \ldots & \ldots     & 1.20 & 1.20$\pm$0.01 \\
SCR~0844-0637   & 2014-02-18 & -0.30 &                & \ldots &            & 1.19 & \\
LP~728-71       & 2014-03-08 & -0.04 & -0.05$\pm$0.01 & \ldots & \ldots     & 1.23 & 1.21$\pm$0.02 \\
LP~728-71       & 2014-03-08 & -0.06 &                & \ldots &            & 1.20 & \\
SCR~1012-3124AB & 2014-03-15 & -4.37 & -4.39$\pm$0.02 & 329    & 334$\pm$5  & 0.61 & 0.63$\pm$0.02 \\
SCR~1012-3124AB & 2014-03-15 & -4.41 &                & 339    &            & 0.64 & \\
TWA~3ABCD       & 2014-02-17 &-17.74 &-17.66$\pm$0.08 & 281    & 278$\pm$3  & 0.73 & 0.73$\pm$0.01 \\
TWA~3ABCD       & 2014-02-17 &-17.58 &                & 275    &            & 0.73 & \\
SCR~1121-3845   & 2014-02-15 & -3.46 & -3.44$\pm$0.02 & 229    & 226$\pm$3  & 0.76 & 0.76$\pm$0.01 \\
SCR~1121-3845   & 2014-02-15 & -3.42 &                & 223    &            & 0.77 & \\
TWA~5ABC        & 2014-02-27 & -4.20 & -4.13$\pm$0.09 & 340    & 335$\pm$6  & 0.79 & 0.80$\pm$0.01 \\
TWA~5ABC        & 2014-02-27 & -4.01 &                & 327    &            & 0.82 & \\
RX~1132-3019    & 2014-01-30 & -5.50 & -5.43$\pm$0.07 & 409    & 396$\pm$12 & 1.51 & 1.34$\pm$0.16 \\
RX~1132-3019    & 2014-01-30 & -5.37 &                & 384    &            & 1.18 & \\
RX~1132-2651A   & 2014-02-27 & -7.09 & -7.12$\pm$0.03 & 278    & 283$\pm$5  & 0.87 & 0.87$\pm$0.01 \\
RX~1132-2651A   & 2014-02-27 & -7.16 &                & 289    &            & 0.87 & \\
SIPS~1145-4055  & 2013-05-25 & -1.48 & -1.42$\pm$0.09 & \ldots & \ldots     & 1.60 & 1.61$\pm$0.02 \\
SIPS~1145-4055  & 2013-05-25 & -1.27 &                & \ldots &            & 1.65 & \\
LP~851-410      & 2014-02-02 & -2.03 & -0.86$\pm$1.04 & \ldots & \ldots     & 0.49 & 0.50$\pm$0.01 \\
LP~851-410      & 2014-02-02 & +0.06 &                & \ldots &            & 0.50 & \\
SCR~1200-1731   & 2014-01-30 & -6.47 & -6.34$\pm$0.11 & 231    & 266$\pm$31 & 0.64 & 0.60$\pm$0.03 \\
SCR~1200-1731   & 2014-01-30 & -6.25 &                & 294    &            & 0.58 & \\
2MASS~1207-3247 & 2014-02-27 & -2.83 & -2.81$\pm$0.05 &  93    &  64$\pm$37 & 0.62 & 0.62$\pm$0.03 \\
2MASS~1207-3247 & 2014-02-27 & -2.79 &                &  22    &            & 0.66 & \\
2MASS~1207-3247 & 2014-02-27 & -3.95 &                & -91    &            & 0.41 & \\
2MASS~1207-3247 & 2014-02-27 & -2.82 &                &  96    &            & 0.60 & \\
L~758-107       & 2014-01-19 & +0.06 & +0.06$\pm$0.01 & \ldots & \ldots     & 1.02 & 1.01$\pm$0.01 \\
L~758-107       & 2014-01-19 & +0.05 &                & \ldots &            & 1.01 & \\
SCR~1230-3300   & 2014-01-26 & -0.33 & -0.31$\pm$0.01 & \ldots & \ldots     & 0.88 & 0.89$\pm$0.01 \\
SCR~1230-3300   & 2014-01-26 & -0.30 &                & \ldots &            & 0.89 & \\
SCR~1233-3641   & 2014-02-02 & -3.62 & -3.59$\pm$0.03 & \ldots & \ldots     & 0.50 & 0.83$\pm$0.33 \\
SCR~1233-3641   & 2014-02-02 & -3.56 &                & \ldots &            & 1.15 & \\
SCR~1237-4021   & 2014-01-21 & -4.39 & -4.41$\pm$0.02 & 234    & 232$\pm$3  & 0.69 & 0.69$\pm$0.01 \\
SCR~1237-4021   & 2014-01-21 & -4.43 &                & 229    &            & 0.70 & \\
SCR~1238-2703   & 2014-01-14 & -2.20 & -2.21$\pm$0.01 & \ldots & \ldots     & 1.12 & 1.11$\pm$0.01 \\
SCR~1238-2703   & 2014-01-14 & -2.22 &                & \ldots &            & 1.10 & \\
SCR~1316-0858   & 2014-03-16 & -2.84 & -2.87$\pm$0.03 & \ldots & \ldots     & 1.78 & 1.79$\pm$0.01 \\
SCR~1316-0858   & 2014-03-16 & -2.90 &                & \ldots &            & 1.80 & \\
SCR~1321-1052   & 2014-04-06 & -5.61 & -5.72$\pm$0.11 & \ldots & \ldots     & 0.86 & 0.75$\pm$0.11 \\
SCR~1321-1052   & 2014-04-06 & -5.84 &                & \ldots &            & 0.64 & \\
SCR~1421-0916   & 2014-03-17 & +0.08 & +0.06$\pm$0.01 & \ldots & \ldots     & 0.93 & 0.93$\pm$0.01 \\
SCR~1421-0916   & 2014-03-17 & +0.06 &                & \ldots &            & 0.93 & \\
SCR~1421-0755   & 2014-03-15 & -0.01 & -0.01$\pm$0.02 & \ldots & \ldots     & 0.91 & 0.90$\pm$0.01 \\
SCR~1421-0755   & 2014-03-15 & +0.02 &                & \ldots &            & 0.88 & \\
SCR~1425-4113AB & 2013-04-30 & -4.76 & -4.72$\pm$0.05 & 245    & 244$\pm$1  & 0.73 & 0.76$\pm$0.03 \\
SCR~1425-4113AB & 2013-04-30 & -4.66 &                & 242    &            & 0.79 & \\
SCR~1438-3941   & 2013-04-30 & -0.12 & -0.10$\pm$0.01 & \ldots & \ldots     & 0.78 & 0.78$\pm$0.01 \\
SCR~1438-3941   & 2013-04-30 & -0.09 &                & \ldots &            & 0.79 & \\
LP~914-6        & 2013-05-01 & -0.18 & -0.17$\pm$0.01 & \ldots & \ldots     & 1.24 & 1.23$\pm$0.01 \\
LP~914-6        & 2013-05-01 & -0.17 &                & \ldots &            & 1.23 & \\
SCR~1521-2514   & 2013-05-21 & -3.86 & -3.84$\pm$0.02 & \ldots & \ldots     & 1.06 & 1.06$\pm$0.01 \\
SCR~1521-2514   & 2013-05-21 & -3.82 &                & \ldots &            & 1.06 & \\
SCR~1708-6936   & 2013-04-28 & -4.52 & -4.57$\pm$0.05 & \ldots & \ldots     & 1.19 & 1.16$\pm$0.04 \\
SCR~1708-6936   & 2013-04-28 & -4.62 &                & \ldots &            & 1.12 & \\
SCR~1816-6305   & 2013-04-28 & -0.02 & -0.11$\pm$0.08 & \ldots & \ldots     & 0.85 & 0.91$\pm$0.05 \\
SCR~1816-6305   & 2013-04-28 & -0.20 &                & \ldots &            & 0.96 & \\
SCR~1816-6305   & 2013-04-28 & -0.10 &                & \ldots &            & 0.92 & \\
SCR~1842-5554A  & 2013-08-20 & -3.55 & -5.54$\pm$1.06 & \ldots & \ldots     & 0.86 & 0.97$\pm$0.03 \\
SCR~1842-5554A  & 2013-08-20 & -3.58 &                & \ldots &            & 1.02 & \\
SCR~1842-5554A  & 2014-03-16 & -6.38 &                & \ldots &            & 0.97 & \\
SCR~1842-5554A  & 2014-03-16 & -5.91 &                & \ldots &            & 0.98 & \\
NLTT~47004AB    & 2013-06-18 & -0.11 & -0.11$\pm$0.01 & \ldots & \ldots     & 0.88 & 0.88$\pm$0.01 \\
NLTT~47004AB    & 2013-06-18 & -0.10 &                & \ldots &            & 0.87 & \\
SCR~1856-6922   & 2013-04-28 & -0.10 & -0.08$\pm$0.02 & \ldots & \ldots     & 1.02 & 1.03$\pm$0.01 \\
SCR~1856-6922   & 2013-04-28 & -0.06 &                & \ldots &            & 1.04 & \\
WT~625          & 2013-07-22 & -3.98 & -3.97$\pm$0.01 & \ldots & \ldots     & 1.07 & 1.06$\pm$0.01 \\
WT~625          & 2013-07-22 & -3.96 &                & \ldots &            & 1.06 & \\
SCR~1922-6310   & 2013-05-18 & -4.83 & -4.79$\pm$0.04 & \ldots & \ldots     & 1.14 & 1.14$\pm$0.01 \\
SCR~1922-6310   & 2013-05-18 & -4.75 &                & \ldots &            & 1.15 & \\
RX~1924-3442    & 2013-08-20 &-10.13 &-10.18$\pm$0.05 & \ldots & \ldots     & 1.08 & 1.17$\pm$0.09 \\
RX~1924-3442    & 2013-08-20 &-10.23 &                & \ldots &            & 1.27 & \\
SCR~1926-5331   & 2013-08-20 & -2.33 & -2.27$\pm$0.06 & \ldots & \ldots     & 0.86 & 0.83$\pm$0.02 \\
SCR~1926-5331   & 2013-08-20 & -2.21 &                & \ldots &            & 0.81 & \\
SCR~1938-2416   & 2013-06-17 & -0.46 & -0.45$\pm$0.01 & \ldots & \ldots     & 1.06 & 1.05$\pm$0.02 \\
SCR~1938-2416   & 2013-06-17 & -0.44 &                & \ldots &            & 1.03 & \\
SCR~1951-4025   & 2013-06-16 & -0.21 & -0.27$\pm$0.06 & \ldots & \ldots     & 1.00 & 1.02$\pm$0.02 \\
SCR~1951-4025   & 2013-06-16 & -0.33 &                & \ldots &            & 1.04 & \\
SCR~2004-6725A  & 2013-04-28 & -3.50 & -3.55$\pm$0.04 & \ldots & \ldots     & 0.91 & 0.91$\pm$0.01 \\
SCR~2004-6725A  & 2013-04-28 & -3.58 &                & \ldots &            & 0.91 & \\
2MASS~2004-3356 & 2014-04-10 & -8.50 & -8.20$\pm$0.15 & \ldots & \ldots     & 1.40 & 1.26$\pm$0.09 \\
2MASS~2004-3356 & 2014-04-11 & -8.16 &                & \ldots &            & 1.30 & \\
2MASS~2004-3356 & 2014-04-11 & -8.09 &                & \ldots &            & 1.16 & \\
SCR~2008-3519   & 2013-06-16 & -5.45 & -5.42$\pm$0.03 & \ldots & \ldots     & 0.98 & 0.99$\pm$0.01 \\
SCR~2008-3519   & 2013-06-16 & -5.39 &                & \ldots &            & 1.00 & \\
SCR~2010-2801AB & 2013-05-23 & -7.93 & -7.94$\pm$0.02 & \ldots & \ldots     & 0.92 & 0.93$\pm$0.01 \\
SCR~2010-2801AB & 2013-05-23 & -7.97 &                & \ldots &            & 0.94 & \\
L~755-19        & 2013-06-12 & -5.32 & -5.35$\pm$0.03 & \ldots & \ldots     & 1.29 & 1.29$\pm$0.01 \\
L~755-19        & 2013-06-12 & -5.38 &                & \ldots &            & 1.29 & \\
SCR~2107-7056   & 2014-07-15 & -5.29 & -5.31$\pm$0.02 & \ldots & \ldots     & 1.18 & 1.18$\pm$0.01 \\
SCR~2107-7056   & 2014-07-15 & -5.34 &                & \ldots &            & 1.18 & \\
SCR~2107-1304   & 2013-08-22 & -4.27 & -4.21$\pm$0.06 & \ldots & \ldots     & 1.19 & 1.18$\pm$0.01 \\
SCR~2107-1304   & 2013-08-22 & -4.15 &                & \ldots &            & 1.16 & \\
LEHPM~1-4147    & 2013-04-28 & +0.02 & +0.01$\pm$0.01 & \ldots & \ldots     & 0.86 & 0.86$\pm$0.01 \\
LEHPM~1-4147    & 2013-04-28 & -0.01 &                & \ldots &            & 0.87 & \\
SCR~2204-0711   & 2013-08-23 & -0.34 & -0.33$\pm$0.01 & \ldots & \ldots     &-0.09 &-0.07$\pm$0.01 \\
SCR~2204-0711   & 2013-08-23 & -0.32 &                & \ldots &            &-0.06 & \\
SCR~2237-2622   & 2013-08-18 & -6.24 & -6.24$\pm$0.05 & \ldots & \ldots     & 1.46 & 1.46$\pm$0.02 \\
SIPS~2258-1104  & 2013-09-13 & -1.80 & -1.82$\pm$0.01 & \ldots & \ldots     & 1.16 & 1.16$\pm$0.01 \\
SIPS~2258-1104  & 2013-09-13 & -1.83 &                & \ldots &            & 1.16 & \\
LEHPM~1-5404    & 2013-08-20 & -0.08 & -0.09$\pm$0.01 & \ldots & \ldots     & 0.95 & 0.95$\pm$0.01 \\
LEHPM~1-5404    & 2013-08-20 & -0.09 &                & \ldots &            & 0.94 & \\
SCR~2328-6802   & 2013-08-20 & -3.89 & -3.89$\pm$0.01 & \ldots & \ldots     & 0.92 & 0.92$\pm$0.01 \\
SCR~2328-6802   & 2013-08-20 & -3.89 &                & \ldots &            & 0.93 & \\
LTT~9582        & 2013-08-20 & -3.79 & -3.79$\pm$0.01 & \ldots & \ldots     & 1.30 & 1.29$\pm$0.01 \\
LTT~9582        & 2013-08-20 & -3.79 &                & \ldots &            & 1.28 & \\
G~275-71        & 2013-11-30 & -0.06 & -0.01$\pm$0.02 & \ldots & \ldots     & 0.91 & 0.94$\pm$0.02 \\
G~275-71        & 2013-11-30 & -0.01 &                & \ldots &            & 0.95 & \\
LEHPM~1-6053    & 2013-08-23 & -0.07 & -0.06$\pm$0.01 & \ldots & \ldots     & 0.94 & 0.94$\pm$0.01 \\
LEHPM~1-6053    & 2013-08-23 & -0.04 &                & \ldots &            & 0.94 & \\
\hline
\multicolumn{8}{c}{Flux Standards} \\
\hline
LTT~1020        & 2014-02-05 & +0.03 & +0.03$\pm$0.01 & \ldots & \ldots     &-0.92 &-0.92$\pm$0.02 \\
EG~21           & 2013-08-21 & +0.53 & +0.40$\pm$0.30 & \ldots & \ldots     &-0.05 &-0.03$\pm$0.05 \\
EG~21           & 2013-12-01 & +0.52 &                & \ldots &            &-0.14 & \\
EG~21           & 2014-01-17 & -0.13 &                & \ldots &            & 0.04 & \\
EG~21           & 2014-01-18 & +0.64 &                & \ldots &            & 0.00 & \\
LTT~2415        & 2013-04-30 & -0.78 & -0.78$\pm$0.02 & \ldots & \ldots     &-0.05 &-0.05$\pm$0.02 \\
HILTNER~600     & 2014-03-08 & -0.00 & -0.00$\pm$0.01 & \ldots & \ldots     & 0.08 & 0.08$\pm$0.01 \\
LTT~4364        & 2013-06-18 & +0.01 & +0.02$\pm$0.01 & \ldots & \ldots     & 0.01 &-0.39$\pm$0.51 \\
LTT~4364        & 2013-06-18 & +0.02 &                & \ldots &            &-1.04 & \\
LTT~4364        & 2014-01-14 & +0.01 &                & \ldots &            &-0.02 & \\
HR~5501         & 2014-03-15 & +0.97 & +0.97$\pm$0.01 & \ldots & \ldots     &-0.01 &-0.01$\pm$0.01 \\
LTT~7987        & 2014-04-06 & +0.18 & +0.18$\pm$0.03 & \ldots & \ldots     & 0.04 & 0.04$\pm$0.04 
\enddata



\end{deluxetable*}

\end{document}